\pdfoutput=1

\documentclass[11pt,twoside,a4paper,cmspaper,final,collab]{cms-tdr}
\def\svnVersion{3869eb4-D}\def\svnDate{2019/12/02}\def\cmsCernNoTag{CERN-EP-2019-127}\def\cmsCernDate{\today}\def\cmsMessage{"Published in the European Physical Journal C as \href{http://dx.doi.org/10.1140/epjc/s10052-019-7451-7}{\doi{10.1140/epjc/s10052-019-7451-7}.}"}

\begin{document}\cmsNoteHeader{SMP-16-016}

\hyphenation{had-ron-i-za-tion}
\hyphenation{cal-or-i-me-ter}
\hyphenation{de-vices}
\RCS$Id$
\newlength\cmsFigWidth
\ifthenelse{\boolean{cms@external}}{\setlength\cmsFigWidth{0.49\textwidth}}{\setlength\cmsFigWidth{0.65\textwidth}}
\ifthenelse{\boolean{cms@external}}{\providecommand{\cmsLeft}{top\xspace}}{\providecommand{\cmsLeft}{left\xspace}}
\ifthenelse{\boolean{cms@external}}{\providecommand{\cmsRight}{bottom\xspace}}{\providecommand{\cmsRight}{right\xspace}}
\newlength\cmsFigWidthii
\ifthenelse{\boolean{cms@external}}{\setlength\cmsFigWidthii{0.495\textwidth}}{\setlength\cmsFigWidthii{0.75\textwidth}}
\newlength\cmsFigWidthiii
\ifthenelse{\boolean{cms@external}}{\setlength\cmsFigWidthiii{0.415\textwidth}}{\setlength\cmsFigWidthiii{0.495\textwidth}}

\newcommand{\ptgamma}{\ensuremath{\pt^{\Pgg}}}
\newcommand{\etagamma}{\ensuremath{\eta^{\Pgg}}}
\newcommand{\ptjet}{\ensuremath{\pt^{\text{jet}}}}
\newcommand{\etajet}{\ensuremath{\eta^{\text{jet}}}}
\newcommand{\absetagamma}{\ensuremath{\abs{\eta^{\Pgg}}}}
\newcommand{\absetajet}{\ensuremath{\abs{\eta^{\text{jet}}}}}
\newcommand{\abseta}{\ensuremath{\abs{\eta}}}
\newcommand{\qsq}{\ensuremath{Q^{2}}}
\newcommand{\xt}{\ensuremath{x_\text{T}}}

\cmsNoteHeader{SMP-15-008}

\title{Measurements of triple-differential cross sections for inclusive isolated-photon+jet events in $\Pp\Pp$ collisions at $\sqrt{s} = 8\TeV$}
\titlerunning{Differential cross sections for isolated-photon+jet events at $\sqrt{s}=8\TeV$}

\date{\today}

\abstract{
    Measurements are presented of the triple-differential cross section for inclusive isolated-photon+jet
    events in $\Pp\Pp$ collisions at $\sqrt{s} = 8\TeV$ as a function of
    photon transverse momentum ($\ptgamma$), photon pseudorapidity
    ($\etagamma$), and jet pseudorapidity ($\etajet$).  The data
    correspond to an integrated luminosity of $19.7\fbinv$ that probe a broad range of the available phase space, for
    $\absetagamma<1.44$ and $1.57<\absetagamma<2.50$, $\absetajet<2.5$,
    $40< \ptgamma<1000$\GeV, and jet transverse momentum, $\ptjet$, $>$25\GeV.
    The measurements are compared to next-to-leading order perturbative quantum chromodynamics calculations, which reproduce
    the data within uncertainties.}

\hypersetup{%
pdfauthor={CMS Collaboration},%
pdftitle={Measurements of triple-differential cross sections for inclusive isolated-photon+jet events in pp collisions at sqrt(s) = 8 TeV},%
pdfsubject={CMS},%
pdfkeywords={CMS, physics, QCD}}

\maketitle

\section{Introduction}
Direct photons produced in the hard scattering of
partons in proton-proton collisions are sensitive
probes of the perturbative regime of quantum chromodynamics (pQCD)~\cite{Campbell:2016lzl,Chen:2019zmr}
and provide useful constraints on the parton distribution function (PDF) of gluons~\cite{dEnterria:2012kvo,Carminati:2012mm,Campbell:2018wfu}.
At leading order in pQCD, direct photons are produced mainly through quark-gluon scattering
($\Pq\Pg\to \Pq\Pgg$) with smaller contributions from quark antiquark
annihilation ($\qqbar \to \Pg\Pgg$).  Photons can
also be produced via fragmentation of the final state partons.  These latter photons
are typically accompanied by other partons, and their contributions can be experimentally
suppressed by requiring the photons to be isolated from other energy depositions in the calorimeters.
A good understanding of isolated photon production also indirectly impacts all jet measurements at
the LHC, because photon+jet events are commonly used to determine the absolute jet energy-scale.
This process also constitutes a main background in important standard model (SM)
processes, such as $\HGG$, as well as in searches for physics beyond the SM.

This paper presents measurements of the triple-differential inclusive isolated-photon+jet
cross sections using data collected by the CMS
experiment during the 2012 run at $\sqrt{s} = 8\TeV$ corresponding to an integrated
luminosity of $19.7\fbinv$.  Measurement of the cross section
as a function of different combinations of photon and jet pseudorapidities in
the range of $\abseta<2.5$ allows for the
exploration of parton collisions at different values of momentum transfer squared (\qsq) and parton momentum
fraction ($x$).  Given the photon transverse momentum range of $\ptgamma = 40$--1000\GeV, the measurement
probes $\qsq = (\ptgamma)^{2}$ in the range $10^{3}$--$10^{6}\GeV^{2}$,
and $\xt={2\ptgamma}/{\sqrt{s}}$ in the range 0.01--0.25, where
$\xt$ is an approximation to the parton momentum fraction when both photon
and jet are produced centrally.  This measurement is complementary to previous
ones~\cite{Aad:2016xcr,Aad:2013gaa,Chatrchyan:2013mwa,D0:2013lra,ATLAS:2012ar,Abazov:2008er}
in the coverage of the $\qsq-x$ phase space.
The cross section can be written as:
\begin{equation}
\left(\frac{\rd^{3}\sigma}{\rd\ptgamma\rd\absetagamma\rd\absetajet}\right)_{i}=
\frac{1}{{\Delta \ptgamma}_{i}{\Delta \absetagamma}_{i}{\Delta \absetajet}_{i}}{\sum}_{j} U_{ij} \frac{N_{i} p_{i}}{\epsilon_{i} {\lumi^{'}}_{i}},
\end{equation}
where $N_i$ is the number of candidate events, $p_i$ is the signal purity,
$\epsilon_i$ is the detection efficiency, ${\lumi^{'}}_i$ is the effective integrated luminosity,
and ${\Delta \ptgamma}_i$, ${\Delta \absetagamma}_i$, and ${\Delta \absetajet}_i$ are
the bin size in \ptgamma, $ \absetagamma$, and $\absetajet$ in the
$i$th data bin.  $U_{ij}$ is the coefficient of the unfolding matrix between the
true quantity in bin $j$ and measured quantities in bin $i$.

The paper is organized as follows.
Section~\ref{detector} provides a brief introduction to the CMS detector.  Selection and reconstruction of
events, with attention focused on issues of triggering,
photon reconstruction, selections and efficiency, are detailed in Section~\ref{evtsel}. Section~\ref{measurement}
describes the extraction of the signal photons from the energy depositions that originate from
neutral meson decays, the unfolding, and the measurement of differential
cross sections. The results of the measurement, along with comparison with theoretical predictions,
are reported in Section~\ref{theory}. Finally, the summary is presented in Section~\ref{summary}.

\section{The CMS detector}\label{detector}
A detailed description of the CMS
detector, together with definitions of the coordinate system and
relevant kinematic variables, is presented in
Ref.~\cite{Chatrchyan:2008zzk}. The central feature of the CMS
apparatus is a superconducting solenoid of $6\unit{m}$ internal
diameter, providing a magnetic field of $3.8\unit{T}$. Within the field
volume are a silicon pixel and strip tracker, a lead tungstate crystal
electromagnetic calorimeter (ECAL), and a brass and plastic
scintillator hadronic calorimeter (HCAL), each composed of a barrel
and two endcap sections. Muons are measured in gas-ionization
detectors embedded in the steel flux-return yoke outside the
solenoid. Extensive forward calorimetry complements the coverage
provided by the barrel and endcap detectors.

\section{Event reconstruction and selection}\label{evtsel}
The particle-flow algorithm~\cite{1748-0221-12-10-P10003} reconstructs and identifies each
individual particle with an optimized combination of information from
the various elements of the CMS detector.  The identification and
energy measurement of muons, electrons, photons, hadronic jets as well
as the missing transverse momentum come from particle-flow
objects. In addition, the isolations of identified leptons and photons
are measured using the $\pt$ of particle-flow charged hadrons,
 photons, and neutral hadrons.  Jets are reconstructed
using the anti-\kt algorithm with a distance parameter of
$\DR = 0.5$~\cite{Cacciari:2008gp}, where $R$ determines the
size of the jet in $\eta$--$\phi$ space and $\phi$ is measured in radians.
Corrections are applied to the jet energy
as functions of jet $\eta$ and $\pt$ to account for
contributions from additional inelastic proton-proton interactions in the same or
neighboring bunch crossings (pileup),
and for the nonuniform and nonlinear response of the detectors~\cite{JESPaper}.
Jets are further required to have at least minimal energy depositions in
the tracker, HCAL, and ECAL to reject
spurious jets associated with calorimeter noise as well as those associated
with muon and electron candidates that are either mis-reconstructed or isolated~\cite{CMS-PAS-JME-16-003}.
Jets have typical energy resolutions of 15--20\% at 30\GeV, 10\% at 100\GeV, 
and 5\% at 1\TeV~\cite{1748-0221-12-10-P10003}.

Photons are selected from clusters of energy measured in the ECAL with
a small corresponding energy deposition in the HCAL.
For the reconstruction of the endcap photons, the depositions of energy
in the preshower detector are also included.  The calorimeter signals
are calibrated and corrected for changes in the detector response over time.
The energy resolution of isolated photons is about 1\%
in the barrel section of the ECAL for unconverted photons (photons that did not
convert to electrons before reaching the ECAL) in the tens
of \GeV energy range. The remaining barrel photons in the similar energy range have a resolution
of about 1.3\% up to a pseudorapidity of $\abseta = 1.0$, rising to about
2.5\% at $\abseta = 1.4$. In the endcaps, the resolution of unconverted
photons is about 2.5\%, while the remaining endcap photons have a
resolution between 3 and 4\%~\cite{CMS:EGM-14-001}.

Muons are identified by tracks in the muon spectrometer matched to tracks
in the silicon tracker. Quality requirements are placed on the silicon tracker
and muon spectrometer track measurements as well as on the
matching between them.
Matching muon spectrometer tracks to tracks measured in the
silicon tracker results in a relative $\pt$ resolution
of 1.3--2.0\% for muons in the momentum range $20 <\pt < 100\GeV$ in
the barrel ($\abseta<1.2$) and better than 6\% in the endcaps
($1.2<\abseta<2.4$)~\cite{Chatrchyan:2012xi}.

Events selected for this analysis are recorded using a two-level
trigger system~\cite{1748-0221-12-01-P01020}.  A hardware based level-1 trigger requires a cluster
of energy deposited within the ECAL above a pre-defined
$\pt$ threshold.
This threshold is $\pt>20$ or 22\GeV, and is raised to 30\GeV
at high luminosity to keep trigger rates at manageable levels.
The CMS high-level trigger (HLT) applies a
more complicated ECAL energy clustering algorithm than that of level-1, and
requires additional $\pt$ trigger thresholds ranging from 30 to 150\GeV.
HLT triggers with thresholds below 90\GeV have additional loose
calorimetric identification requirements, based on the electromagnetic (EM) shower,
and are prescaled such that only a fraction of events satisfying the trigger
requirements are recorded.
Since the trigger rates for lower $\pt$ threshold triggers are controlled by
applying larger prescale factors, the effective luminosity is smaller for the
lower $\pt$ regions. Triggers are combined for different $\pt$ ranges
to maximize the number of events without loss of efficiency.

Samples of simulated events used for signal and background studies are described below.
Events from both photon+jet production and QCD multijet production with enhanced EM content
are generated using $\PYTHIA$ version 6.426~\cite{1126-6708-2006-05-026}, and passed through the
full CMS detector simulation implemented in $\GEANTfour$~\cite{Agostinelli:2002hh}.
The EM-enriched QCD sample is generated by applying a filter that is designed
to enhance the production efficiency of fake photons from jets with EM
fluctuations. The filter accepts events having photons, electrons, or neutral hadrons with:
(i) a $\pt>15\GeV$ within a small region, and (ii) no more than one charged particle
in a cone of $\DR = \sqrt{\smash[b]{(\Delta\eta)^2+(\Delta\phi)^2}} < 0.2$.  Samples for reconstruction
efficiency studies of inclusive $\PZ/\Pgg^{*} \to \Pe^+\Pe^-$ and $\PZ/\Pgg^{*} \to \Pgm^+\Pgm^-\Pgg$ are generated using $\MADGRAPH$ 5.1.5.11~\cite{1126-6708-2003-02-027}.
For generation purposes, the CTEQ6L~\cite{Pumplin:2002vw} parton distribution functions are
used along with underlying event tune Z2*~\cite{Chatrchyan2013} for all MC samples.
All the samples include simulation of the multiple \Pp\Pp interactions
taking place in each bunch crossing, which are weighted to produce
the pileup distribution observed in data.

Events selected with the single-photon trigger are chosen offline by requiring at
least one photon candidate with $\ptgamma>40$\GeV.  Photon candidates
must either be in the barrel ($\abseta<1.44$) or endcap
($1.57<\absetagamma<2.50$) detector regions.
The leading jet is required to be separated from the photon candidate by $\DR
>0.5$, pass the
jet identification requirements, and have $\ptjet>25$\GeV~and $\abseta<2.5$. 
Therefore, dijet events where a photon is radiated in a parton shower are included. 

The dominant background originates from the
decays of neutral hadrons, such as $\Pgpz$ and $\Pgh$ mesons, into photon
pairs with small angular separation.
To separate signal photons from this background, photons are selected by requiring a
narrow transverse shower shape in the ECAL (in the $\eta$ coordinate),
no matching reconstructed track candidates (except for electron tracks from photon conversion),
and minimal energy measured in the HCAL region matched
to the ECAL shower.  Photon candidates are further required
to be isolated from nearby particle-flow candidates, such as
charged hadrons and photons, after removing those consistent with pileup~\cite{CMS:EGM-14-001}.
A photon candidate is defined as isolated from charged hadrons if
the sum of the $\pt$ of the charged hadron particle-flow candidates
in a cone of radius $\DR<0.3$
around its direction is less than 5\GeV.
To limit correlations of the selected photon candidate's shower energy
with other photon quantities,
an area in the vicinity of the photon candidate is eliminated in the
calculation of the photon isolation (calculated similarly to charged hadron
isolation but from the $\pt$ sum of the photon particle-flow candidates),
leading to smaller correlation overall.
Because of the pileup subtraction, the final photon isolation
may be negative as calculated.  Final photon candidates are
required to have less than 0.0\GeV for $\abseta<1.44$,  $-0.5\GeV$
for $1.5<\abseta<2.1$, and $-1.0\GeV$ for $2.1<\abseta<2.5$.

Several quantities related to the shape of the EM shower
are then used in a boosted-decision-tree (BDT)~\cite{ROE2005577} to discriminate between
direct photons and photons from hadronic activity.
These quantities include the
transverse width of the cluster in the $\eta$ and $\phi$ coordinates in the ECAL, the
calorimetry-based likelihood of this shower to come from a conversion,
the pseudorapidity of the cluster, and the average pileup
energy density of the event.  Simulated samples of photons originating from
photon+jet events, where the reconstructed photons are matched to the
generated photon, are used as training samples for the signal.  Samples of simulated QCD
multijet events selected at generation level as containing
electromagnetically decaying final particles
are used for background training.  
The background contribution from electrons 
misidentified as photons is determined from simulation, 
using $\PW\to\Pe\nu$ sample, and found 
to be many orders of magnitude smaller than the 
QCD multijet background. Therefore, this background is not considered 
in the BDT training.
The output from this BDT is
then used to statistically quantify the fraction of true photons in
the candidate sample.

The efficiency of the photon selection is estimated from simulated photon+jet
events.  To validate the efficiency, large
samples of $\PZ\to \Pe^+\Pe^-$ events in data and simulation are compared.
Since the electrons at CMS are reconstructed by pairing ECAL energy depositions
with the tracks in the tracker, electron showers can be reconstructed as photons
to validate photon selection and identification.
The trigger efficiency is measured to be
approximately 100 (97)\% with an uncertainty of $\approx$3 (2)\% for barrel (endcap) events
above the corresponding trigger
thresholds.  To maintain well-defined trigger efficiencies and
effective luminosities, the bins for the cross section are chosen so that maximum
efficiency is maintained for each trigger with a separate threshold.
The photon selection
efficiencies for the offline preselection and isolation criteria are estimated
to be $84\pm3.4$, $83\pm6.2$, $81\pm6.5$, and $88\pm10.1\%$ in
$\abseta<0.8$, $0.8<\abseta<1.44$, $1.56<\abseta<2.1$, and $2.1<\abseta<2.5$ respectively for all bins in $\ptgamma$.  The statistical
uncertainty in these efficiencies is negligible, and the total
uncertainty is mainly due to differences between the electron and
photon efficiencies observed in the simulation.

\section{Experimental measurement}\label{measurement}
The purity of the selected candidate events is measured bin by bin in photon
$\ptgamma$ and $\etagamma$.  In each bin, a data-based template for the BDT output is
defined for the background, and a simulation-based template is
defined for the signal. The final purity is estimated using a binned maximum likelihood method~\cite{Verkerke:622147}:
\begin{equation}
F(x) = f_{\text{sig}}S(x) + (1-f_{\text{sig}})B(x).
\end{equation}
Here $x$ is the BDT output, $F(x)$ denotes the fit template, $S(x)$ denotes the unity normalized
signal template distribution, and $B(x)$ denotes the unity normalized background template distribution.  The
$f_{\text{sig}}$ parameter describes the signal purity present in the data and is obtained by maximizing the likelihood, which is equivalent to
minimizing the negative of the log-likelihood defined as,
\begin{equation}
-\log{L(f_{\text{sig}};x_1,x_2,\ldots x_N)} = -\Sigma_N\log{F(x_i|f_{\text{sig}})}.
\end{equation}
In the above equation, $L(f_{\text{sig}};x_1,x_2,\ldots x_N)$ is the likelihood function as a function of the
$f_{\text{sig}}$ parameter, $x_i$ represent the individual observed values, and $N$ represents the total number of data points.
The template shape uncertainties are not treated as nuisance parameters, but are characterized using sample
experiments as detailed in Sections~\ref{sigtemplate} and~\ref{bgtemplate} below.

\subsection{Signal templates} \label{sigtemplate}
Signal templates are obtained using photon+jet simulated events.
Because the signal template is obtained from simulation,
a data control sample is used to estimate potential differences between
data and simulation. Samples of $\PZ/\Pgg^{*}\to \Pgm^+\Pgm^-\Pgg$ events
are obtained by selecting events
in which there are two muons and a photon candidate that is produced via final-state
radiation from one of the muons.
Requiring that the dimuon mass be
less than the mass of the on-shell $\PZ$ boson allows for the reconstruction of
a mass peak in the three-body mass ($m_{\Pgm^+\Pgm^-\Pgg}$) distribution.
The sample of events in the peak of the distribution, $80<m_{\Pgm^+\Pgm^-\Pgg}<100$\GeV, is enriched with 
photons, though some background under the peak remains.  The remaining
background in the BDT distribution is estimated using the sidebands, which are obtained 
by inverting the $m_{\Pgm^+\Pgm^-\Pgg}$ criteria, and subtracted.  The resulting distribution for data photons is
then compared to the response in the simulation in the limited range
of $\ptgamma$ available.  The difference is assigned as a
systematic uncertainty in the signal shape for all $\ptgamma$,
in separate bins of $\etagamma$.

\subsection{Background templates} \label{bgtemplate}
The background BDT templates are obtained using a data sideband in pileup-corrected
particle-flow photon isolation.  Except for the photon isolation constraint,
the sideband data is required to pass the same requirements as the signal.
Sideband optimization is performed
using simulations to select a photon isolation region with sufficient
amount of data and minimum correlations between this quantity and the output of
the BDT that is used to fit for the final purity.
Using a mixture of simulated events containing both dijets and photon+jets,
a range of isolation windows are examined.  For each bin of
$\etagamma$ and $\ptgamma$, a range of sideband windows are used
to generate background templates by varying the candidate photon isolation
constraint to an upper bound determined by data set size (nominally 4.5--5\GeV).
Based on the observed data sample size, template shapes are generated randomly
from the simulated shapes and then are used to perform a fit to a separate
mixture of simulation with a known signal fraction.  Based on these
generated shapes, the bias between the known signal
fraction and the signal fraction from the
fit is determined using 500 trials, and the central value of this distribution is taken
as the bias induced by the residual correlations.
Background shapes are estimated separately for the different
pseudorapidity and $\pt$ regions.
The uncertainty in the correction for the bias and the difference
between the final selected data template and the simulated shape are the
systematic uncertainties in the background shape.

\subsection{Fit and systematic uncertainties}
\begin{figure}[htb!]
\centering
\includegraphics[width=\cmsFigWidth]{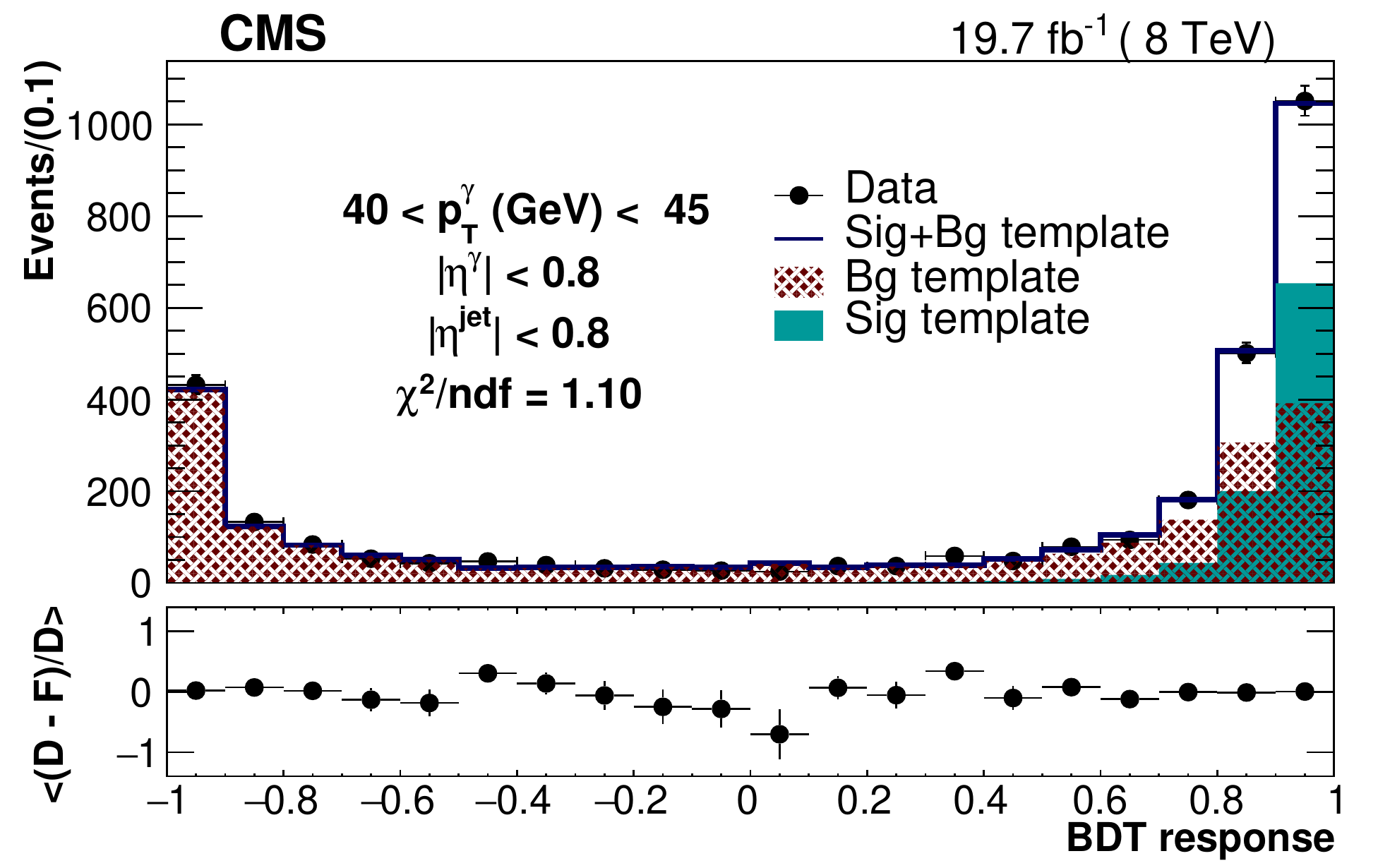}
\caption{ An example fit of candidate boosted-decision-tree distribution with a composite template~(blue histogram). The signal~(background) template is shown by the green~(red) solid~(hatched) region.
The bottom panel shows the mean of the fit values for 500 templates varied within the signal and background shape uncertainties (F) subtracted from data (D) divided by the data.}
\label{fig:PurityFit}
\end{figure}

\begin{figure*}[htb]
\centering
\includegraphics[width=0.49\textwidth]{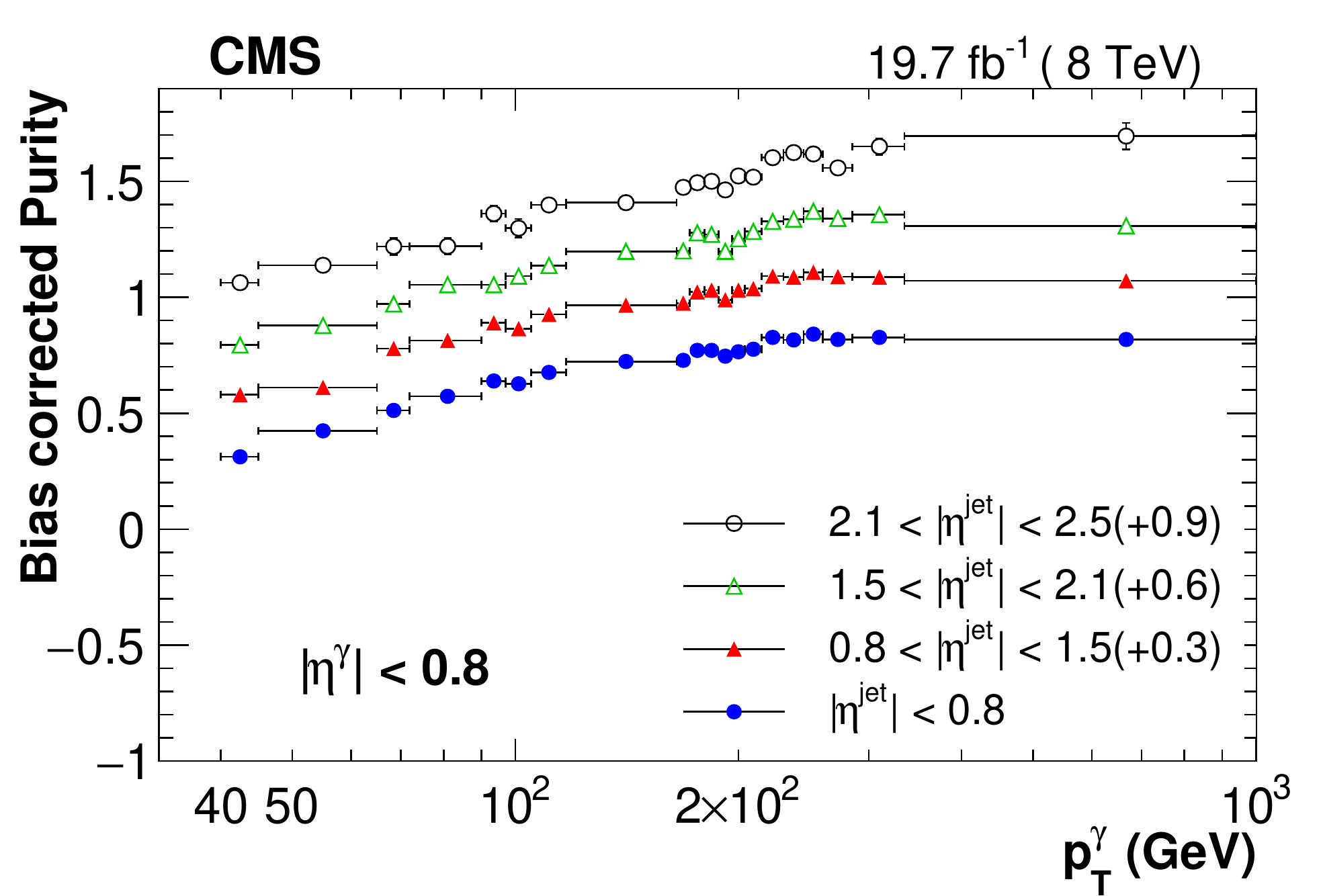}
\includegraphics[width=0.49\textwidth]{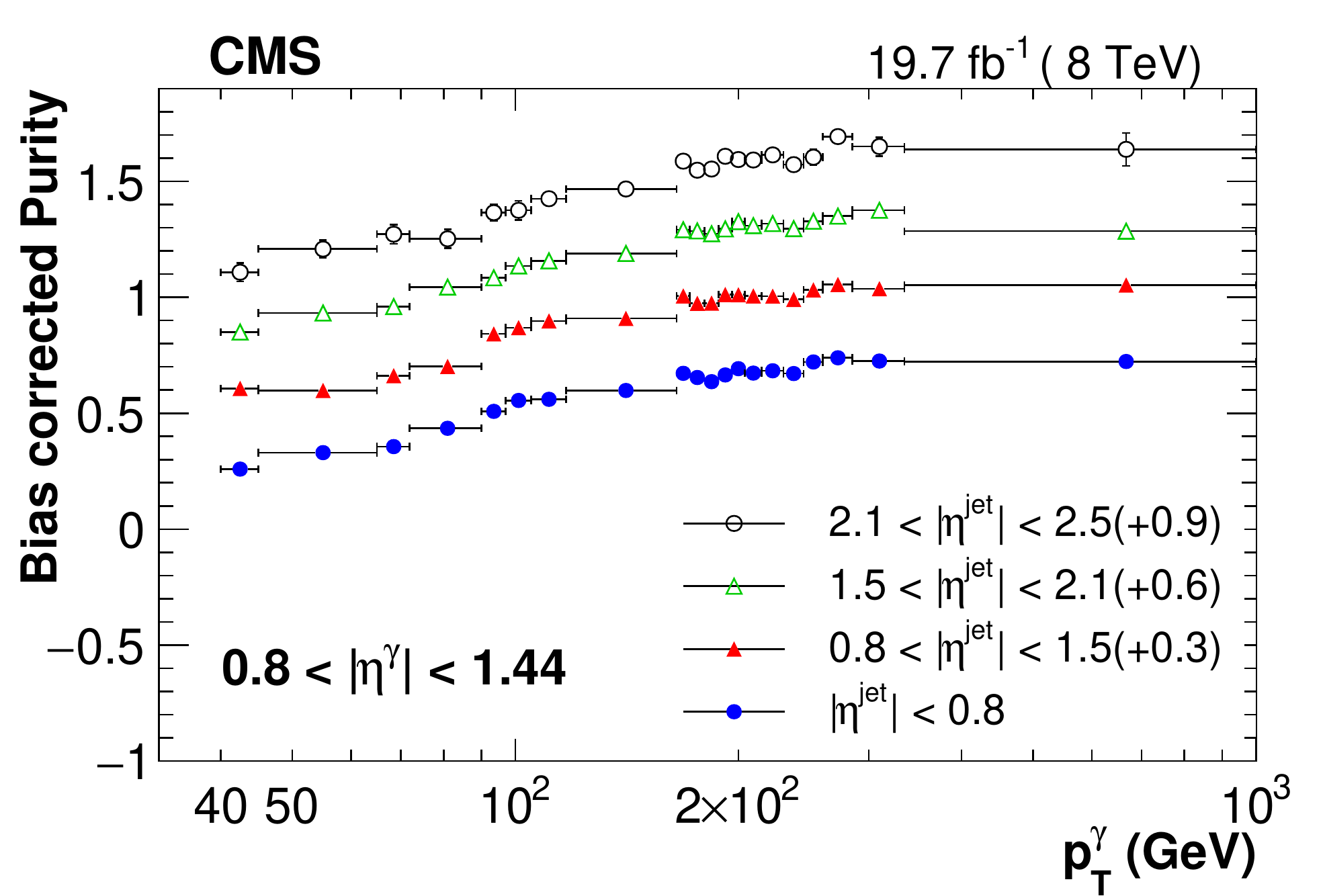}\\
\includegraphics[width=0.49\textwidth]{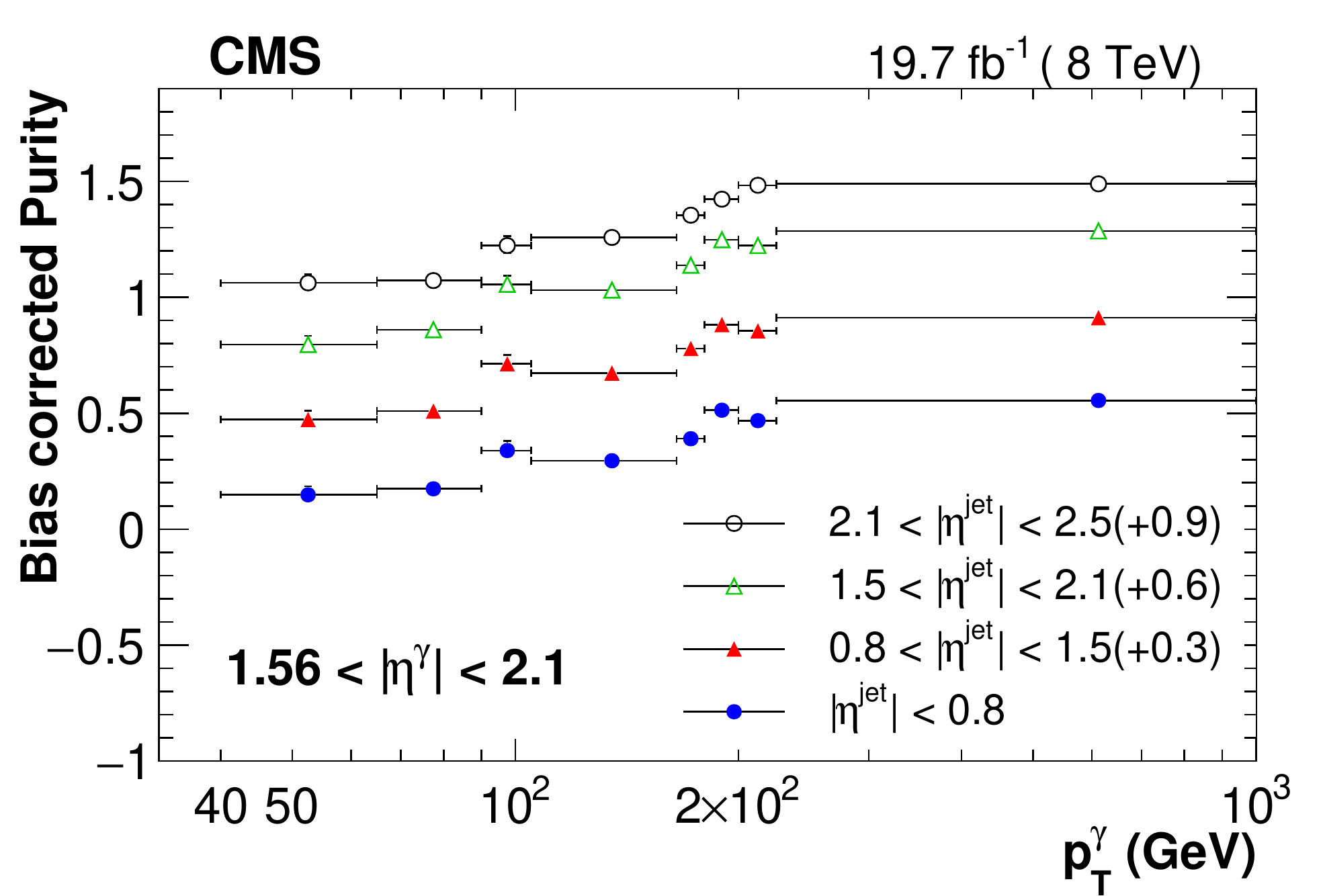}
\includegraphics[width=0.49\textwidth]{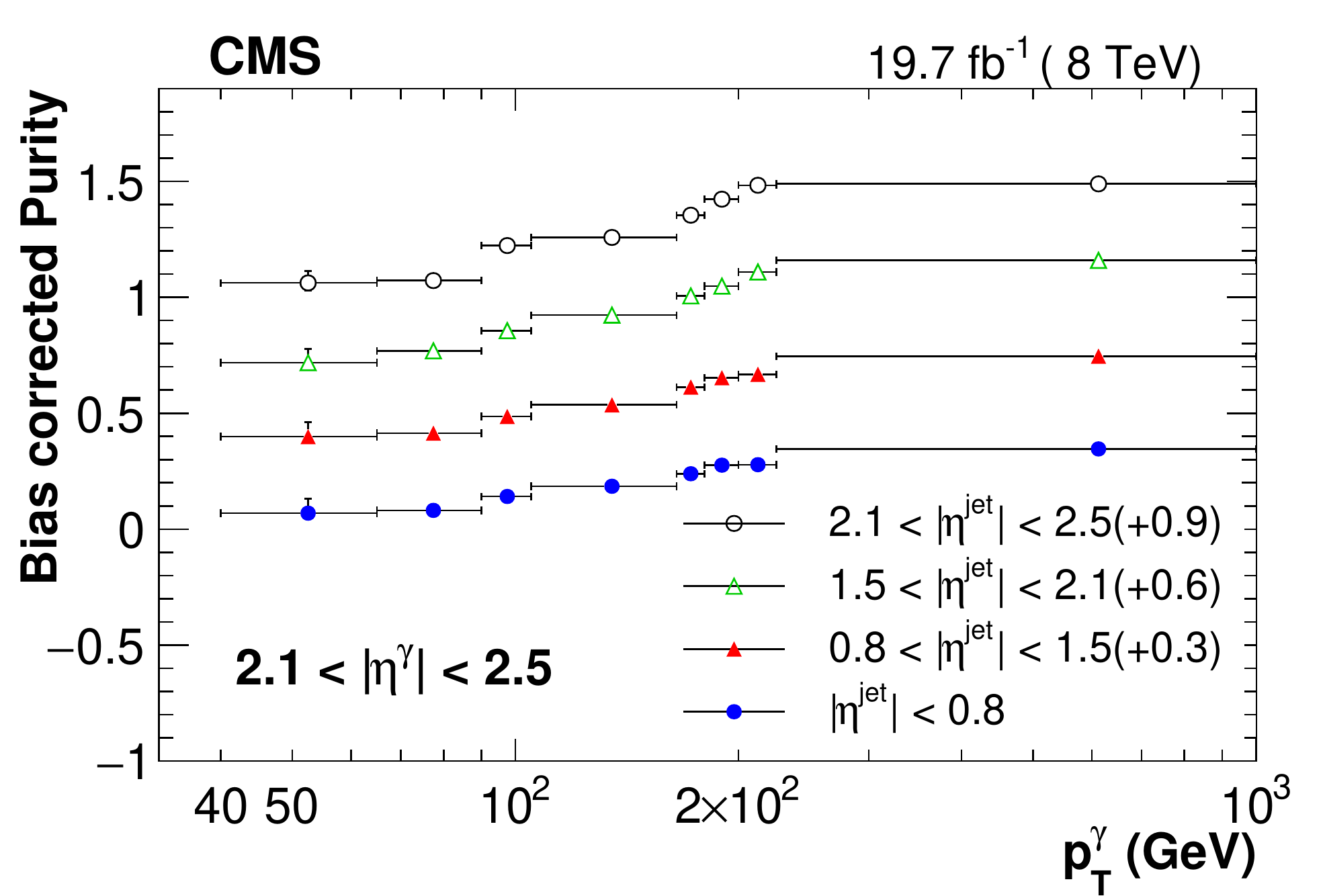}
\caption{ Purity estimates as a function of $\pt^\Pgg$ for different photon and jet pseudorapidity regions. The values are offset by 0.3, 0.6 and 0.9 for $0.8<\absetajet<1.5$, $1.5<\absetajet<2.1$, and $2.1<\absetajet<2.5$ respectively. The total uncertainties are shown as error bars.}
\label{fig:Purity}
\end{figure*}

In each bin of $\absetagamma$, $\absetajet$, and $\ptgamma$ the purity
is estimated by a simultaneous fit to the BDT output using the previously defined signal
and background templates.  An example fit are shown in
Fig.~\ref{fig:PurityFit}.  The uncertainty in this measured purity is
estimated from sample distributions generated by varying the signal and background
fit templates within their respective uncertainties.  For the signal template, where the uncertainty contribution is from
differences between simulation and detector response,
the shapes of sample distributions are obtained by simultaneous
variations across different bins of the BDT template.
On the other hand, the source of background template shape uncertainty is the data sideband statistical uncertainty,
which is uncorrelated across different bins of the BDT distribution.
Therefore, the sample distributions for the background template are created
by allowing the adjacent bins to vary independently of each other.
The purity estimated in each
bin and the associated uncertainty is shown in
Fig.~\ref{fig:Purity}.  The signal purity is lower at larger photon rapidities, where 
the selection criteria are less effective at separating direct photon 
signals from photons from meson decays because of the smaller opening angle between the 
daughter photons. 

The residual bias caused by correlations is minimized, but not completely eliminated, using the sideband optimization process
described in Section~\ref{bgtemplate}.  To compensate for this residual bias, a
correction is applied based on the estimated bias from the simulation.
The correction applied to correct for residual bias in purity
decreases as $\ptgamma$ increases.
These corrections have associated uncertainties from the size of the simulated data samples
and systematic uncertainties of the template shapes. If the bias correction uncertainty
is larger than the associated correction, then the correction is not applied, and the
amount of bias is taken as an additional systematic uncertainty.  The bias-related
uncertainty ranges from 0.01--4.70\% (0.05--10.10\%) in the barrel (endcap) region.
A summary of the uncertainty in the purity from different sources
is provided in Table~\ref{tab:purityun}.
\begin{table*}[htb]
 \centering
\topcaption {Summary of uncertainties in the estimated purity for photons in the barrel (endcap) region.}
  \begin{tabular}{ l  c  c}
 Sources                        &	Barrel photons		&	Endcap photons		\\\hline
 Statistical	                &	0.5--18.7	\%	&	0.8--9.2	\%	\\
 Signal Template Shape	        &	0.2--3.7	\%	&	0.3--7.3	\%	\\
 Background Template Shape	&	0.4--5.2	\%	&	1.3--88.7	\%	\\
 Residual Bias                  &	0.01--4.7	\%	&	0.05--10.1	\%	\\
 Total Systematic 	        &	0.6--7.8	\%	&	1.5--89.3	\%	\\
  \end{tabular}
 \label{tab:purityun}
\end{table*}
\subsection{Unfolding}
The cross section measurements are unfolded within the fiducial volume of acceptance 
and phase space, which are as defined previously in this paper.
With the excellent energy resolution of the ECAL, and the width of the selected bins,
bin-to-bin migrations are small, but still corrected in the final result.
The response matrix is determined from the true generator level $\ptgamma$ and the smeared values obtained from the
simulation.  The D'Agostini iterative unfolding method, implemented in the
RooUnfold~\cite{DAgostini:1995jqe} package, is used to unfold the detector effects.
A systematic
uncertainty in this unfolding, due to the input $\ptgamma$
distribution, is obtained by reweighting the input distribution to
resemble the spectrum observed in data, reproducing the response matrix,
and taking the difference between the unfolded results from the reweighted
response matrix to the unreweighted one.  The final (small) uncertainty from this procedure is
propagated to the final cross section result.

\section{Comparisons with theory}\label{theory}
The measured cross sections are compared with next-to-leading order
(NLO) predictions using the modified version of the
GamJet~\cite{BAER1990127, PhysRevD.42.61} package.
The recent CJ15~\cite{Accardi:2016qay} parton distribution
functions are used as input to this prediction, and uncertainties are assigned based on the deviation
from the 24 pairs of varied PDFs supplied with the CJ15 set.
A tolerance factor of 1, assuming that all of the datasets used in
the PDF calculation are statistically compatible and the experimental
uncertainties are Gaussian, is used for the theoretical prediction.
Set II of Bourhis-Fontannaz-Guillet (BFG)~\cite{Bourhis:1997yu}
fragmentation functions are applied to the matrix element calculations
to estimate the photon production via parton fragmentation.
Although contributions from fragmentation photons are included in these predictions,
an isolation criterion requiring less than 4\GeV of hadronic energy
within a cone of radius $\DR<0.2$ around the photon direction is utilized,
removing a large fraction of them. The central values of
the renormalization, fragmentation, and PDF scales are set to
$\ptgamma$.  The scale uncertainty is quantified by varying each of the scales
by factors of 0.5 and 2.0 independently, and the largest variation
is taken as the systematic uncertainty.  In general, the scale (PDF) uncertainty
is dominant in the low (high) photon pseudorapidity bins, with the total
uncertainty ranging from 10--25\% in most cases, and as high as 70\% in some $\ptgamma$ bins in the
high $\absetajet$ region.

\begin{figure*}[htb!]
\centering
\includegraphics[width=\cmsFigWidthii]{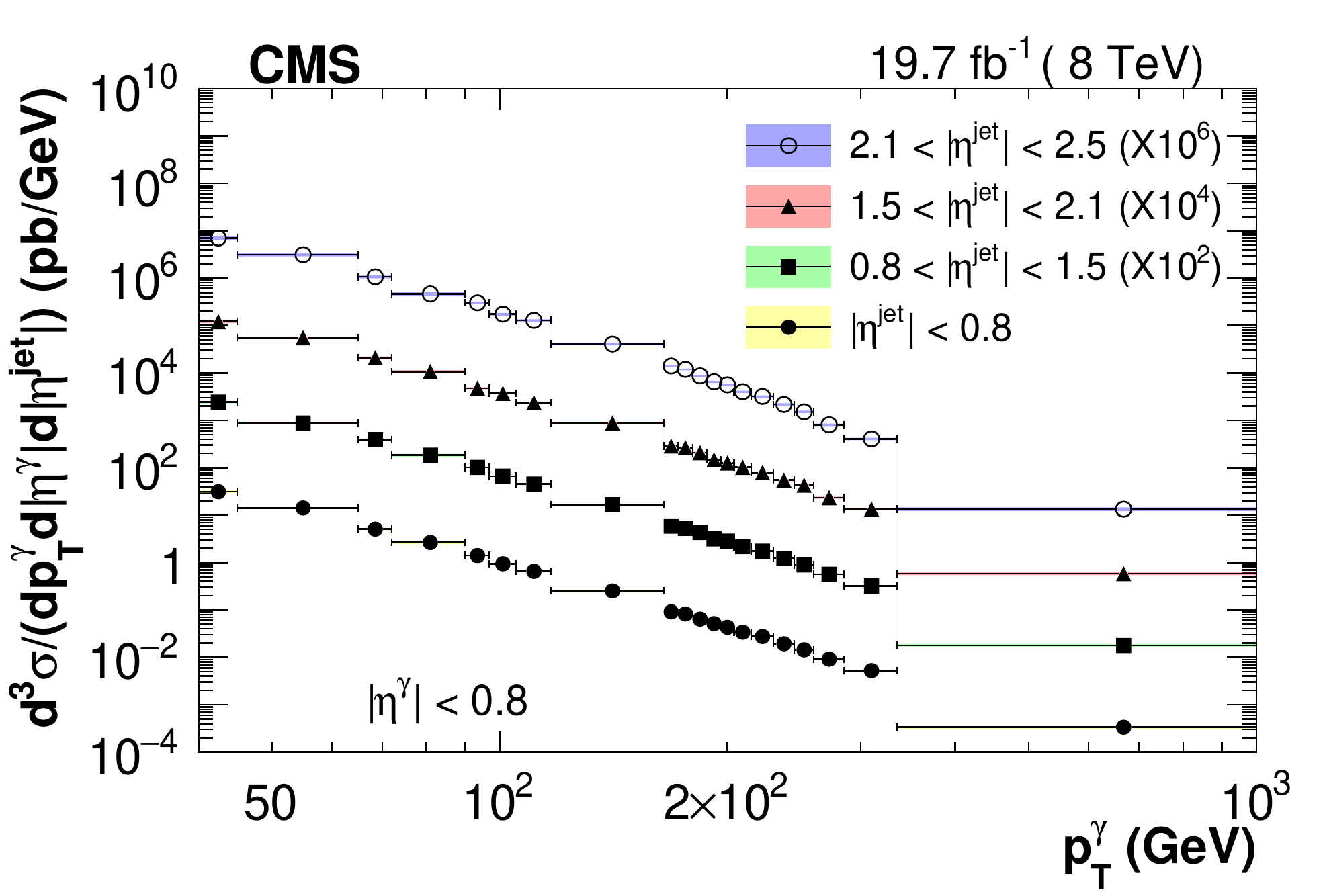}
\includegraphics[width=\cmsFigWidthii]{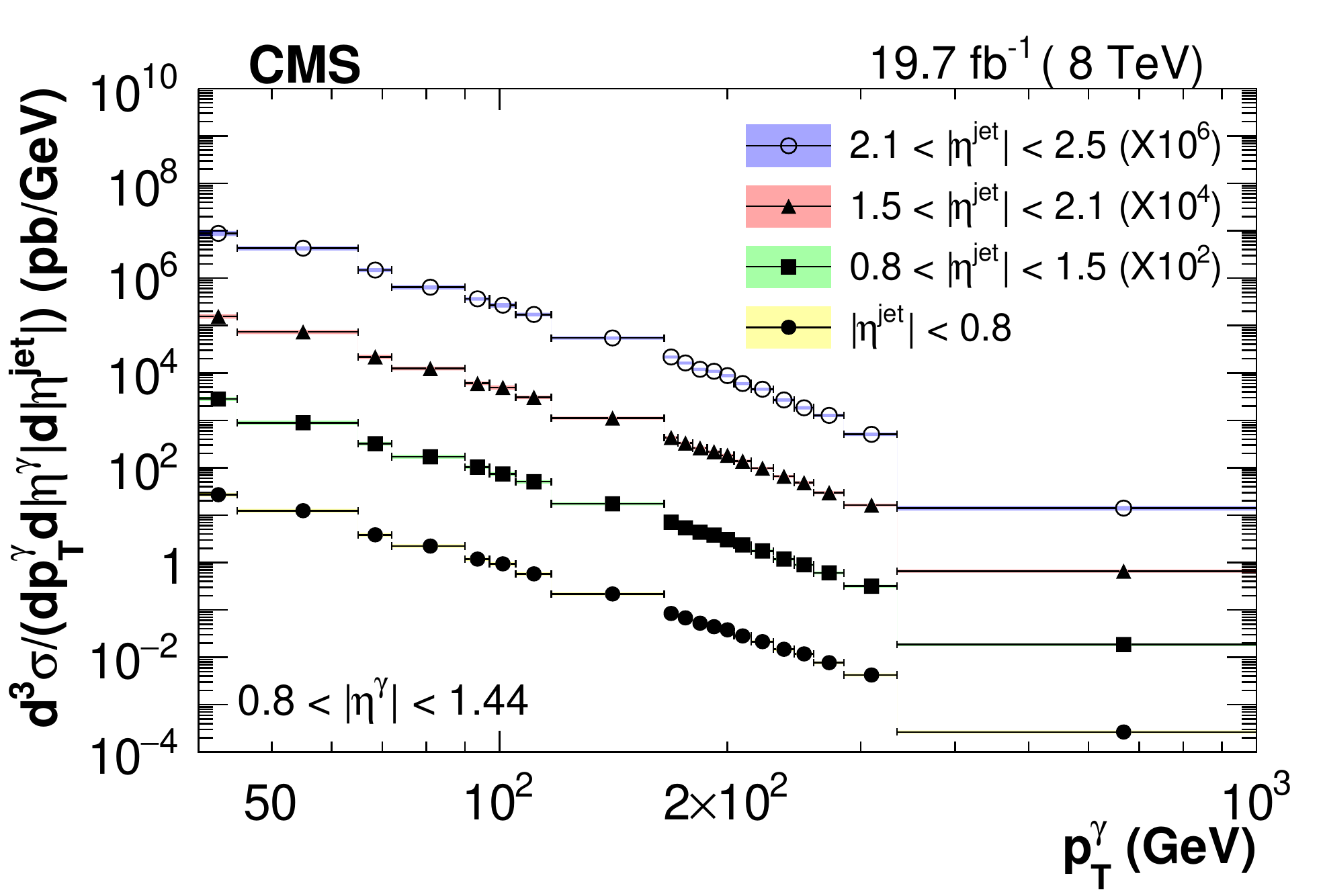}
\caption{Measured triple-differential cross section distributions as a function of $\pt^\Pgg$ in different bins of $\absetajet$ for photons in the barrel region. Note that the distributions are multiplied by a factor of $10^2$, $10^4$ and $10^6$ for $0.8<\absetajet<1.5$, $1.5<\absetajet<2.1$, and $2.1<\absetajet<2.5$ respectively.  The statistical (systematic) uncertainties are shown as error bars (color bands).}
\label{fig:FinalXSecCen}
\end{figure*}

\begin{figure*}[htb!]
\centering
\includegraphics[width=\cmsFigWidthii]{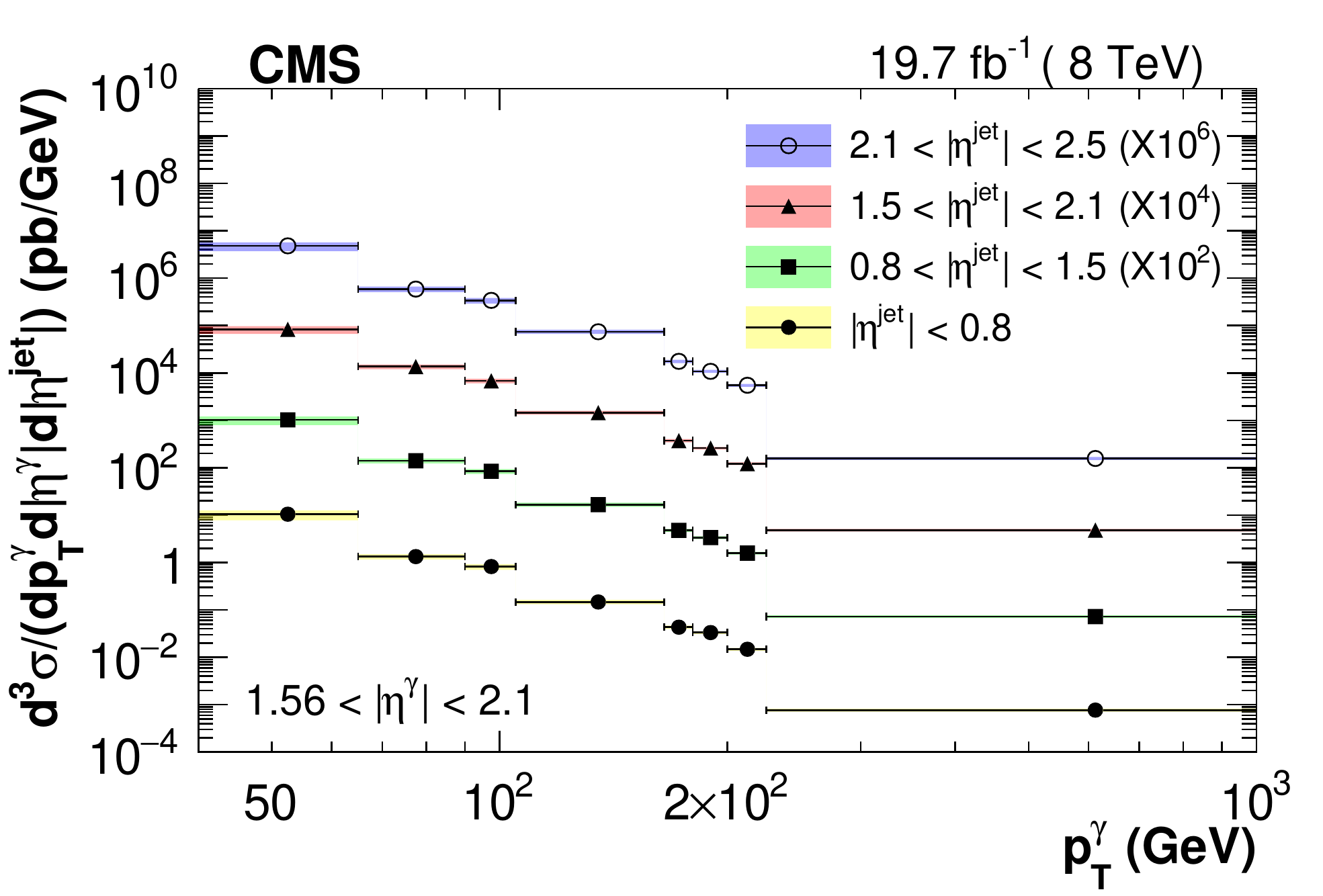}
\includegraphics[width=\cmsFigWidthii]{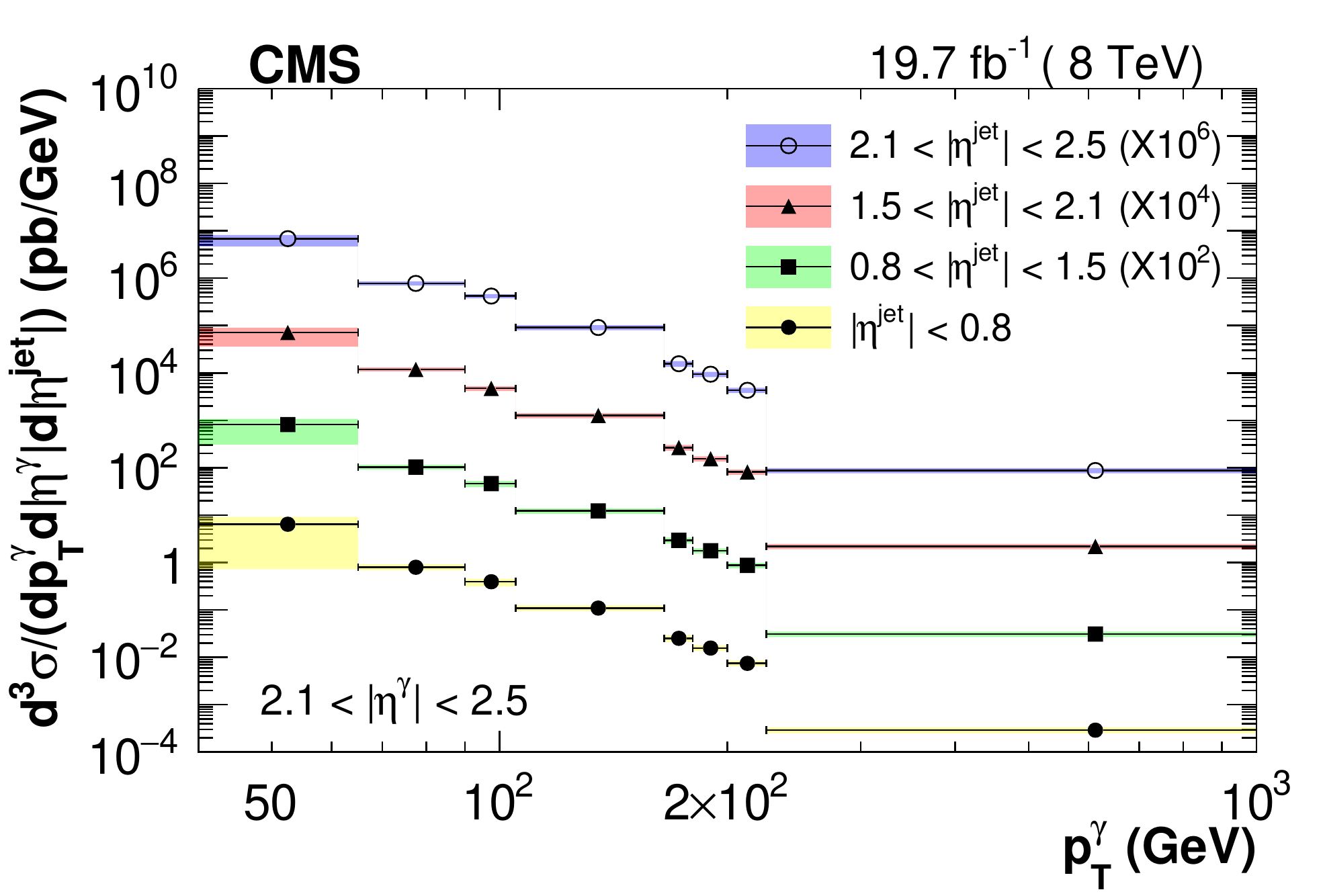}
\caption{Measured triple-differential cross section distributions as a function of $\pt^\Pgg$ in different bins of $\absetajet$ for photons in the endcap region. Note that the distributions are multiplied by a factor of $10^2$, $10^4$ and $10^6$ for $0.8<\absetajet<1.5$, $1.5<\absetajet<2.1$, and $2.1<\absetajet<2.5$ respectively.  The statistical (systematic) uncertainties are shown as error bars (color bands).}
\label{fig:FinalXSecFwd}
\end{figure*}

\begin{figure*}[htb!]
\centering
\includegraphics[width=\cmsFigWidthiii]{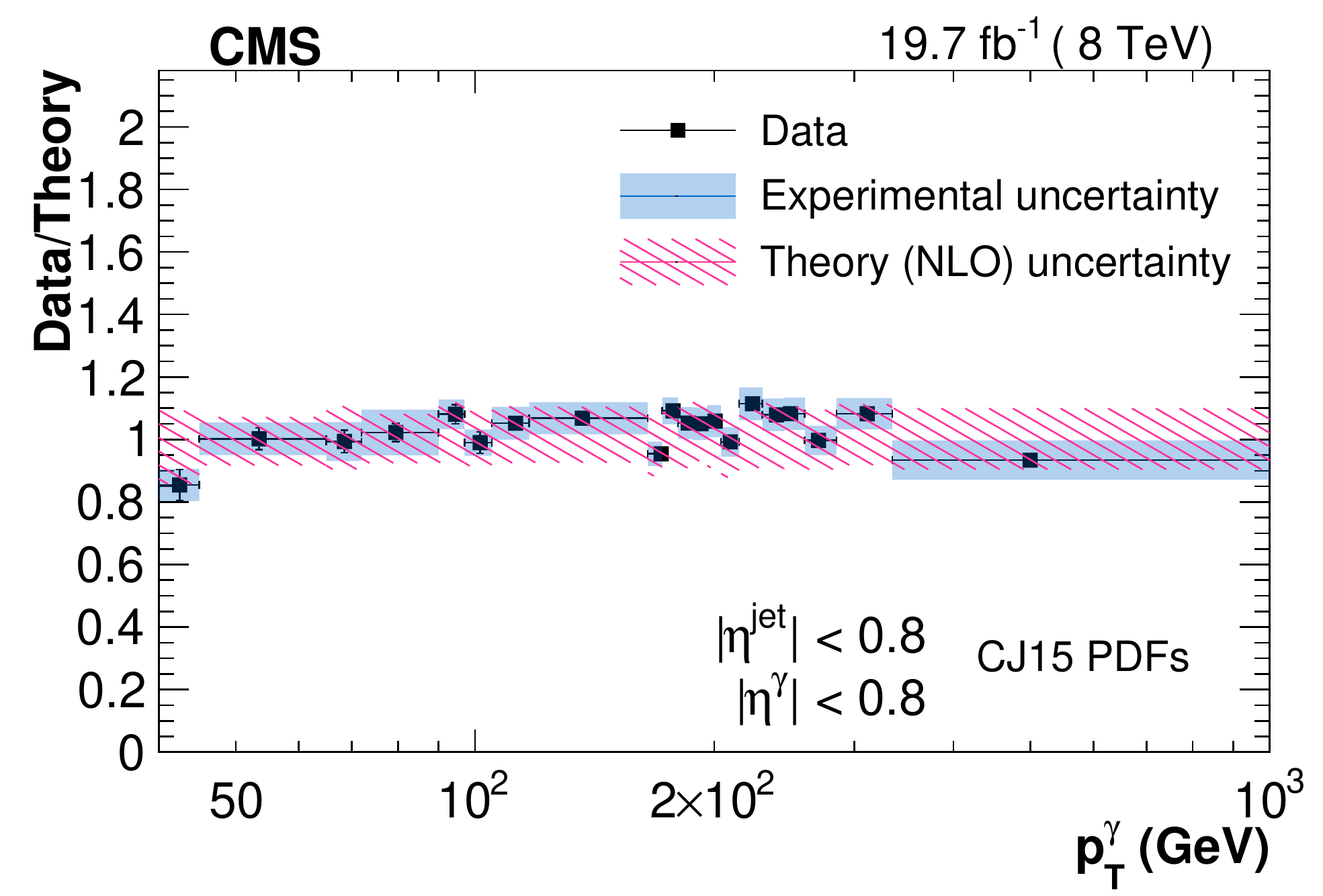}
\includegraphics[width=\cmsFigWidthiii]{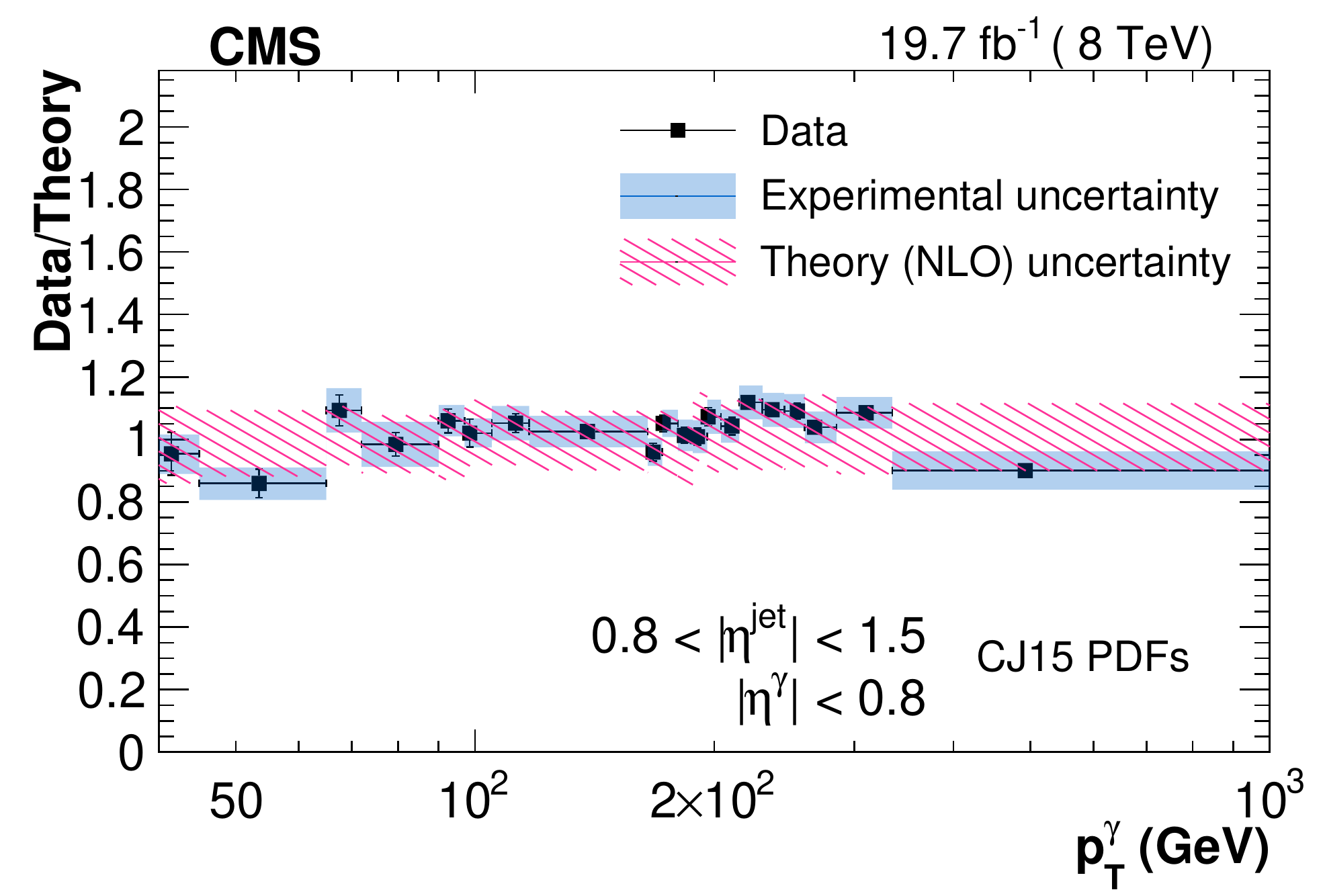}
\includegraphics[width=\cmsFigWidthiii]{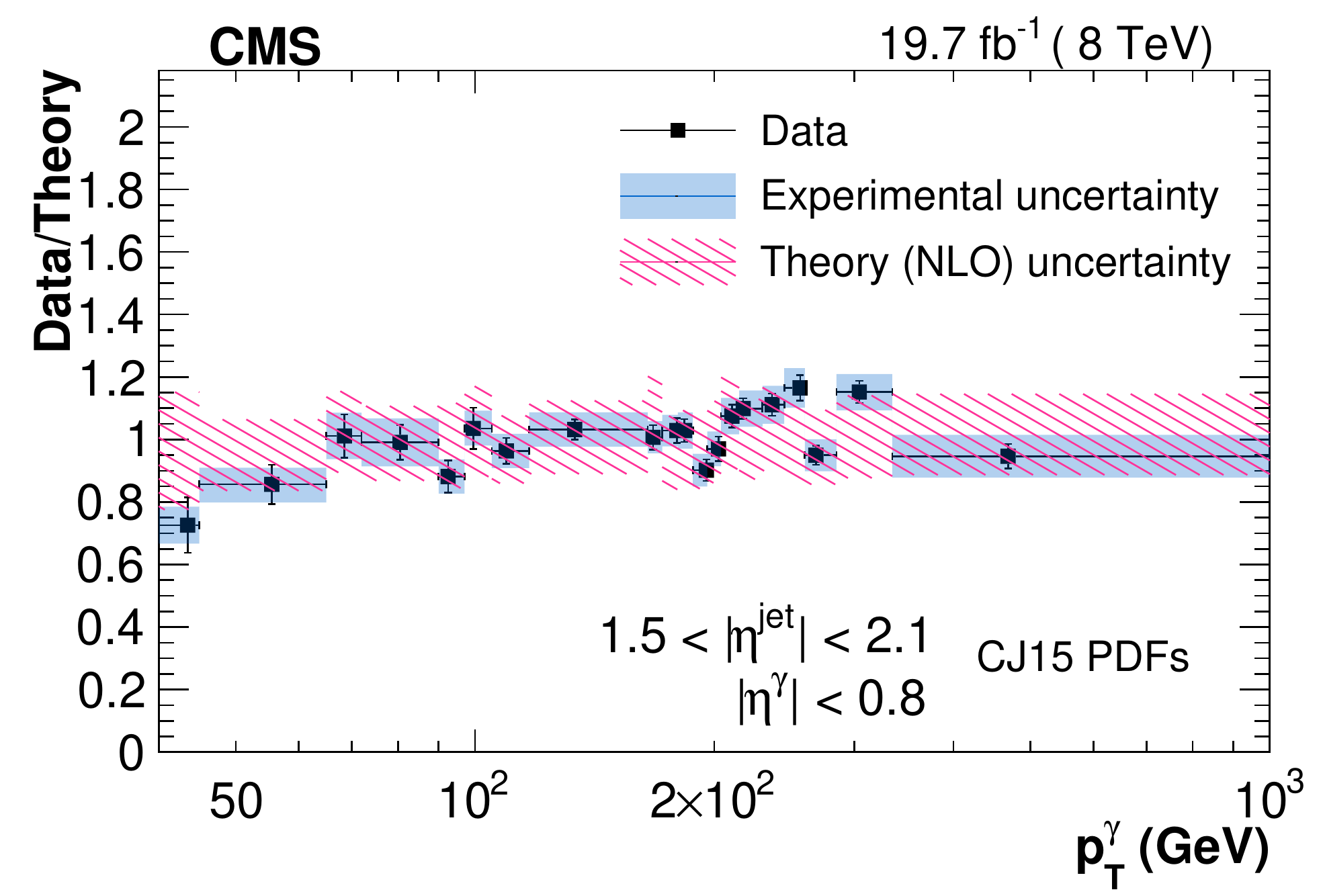}
\includegraphics[width=\cmsFigWidthiii]{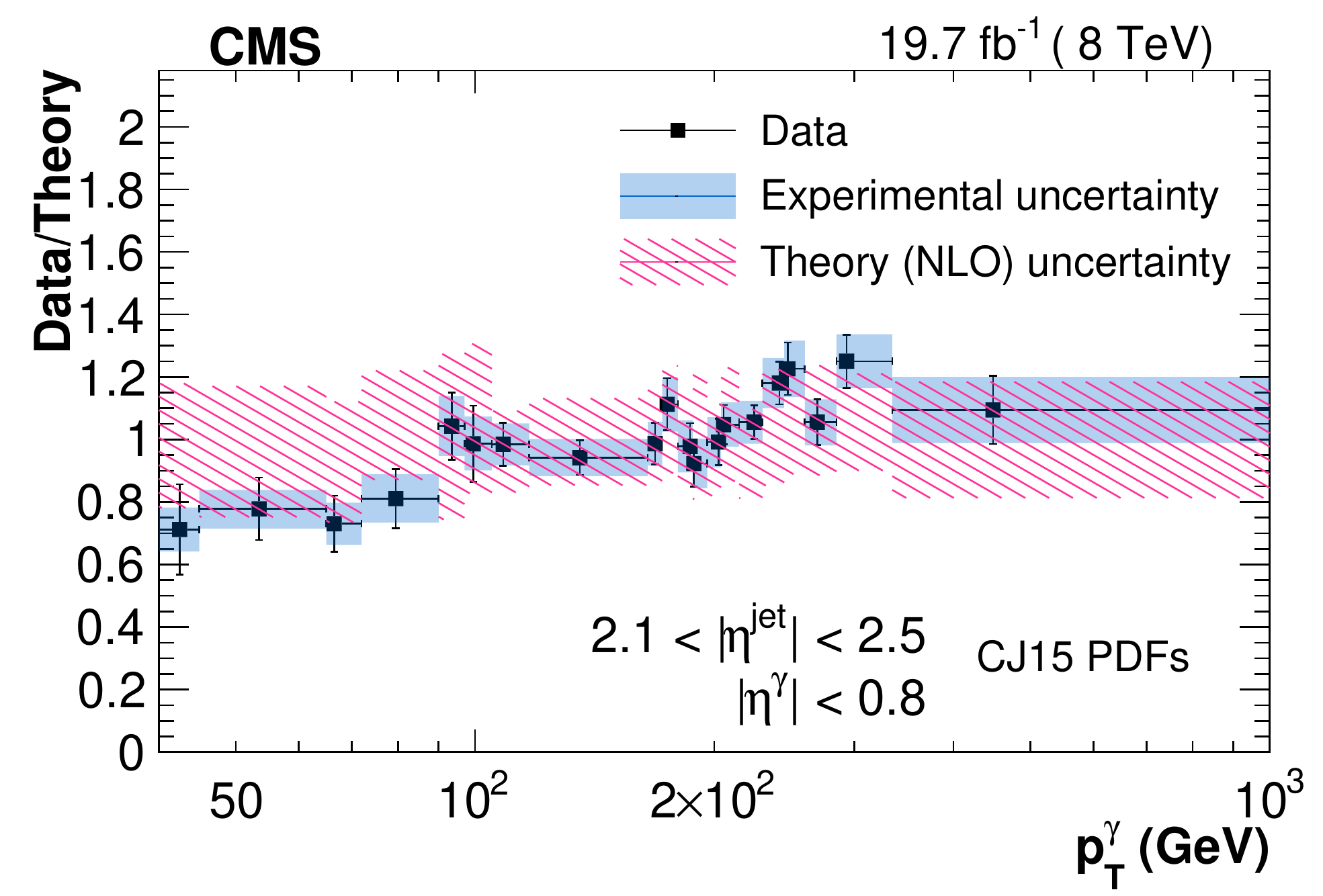}
\caption{Ratio of triple-differential cross sections as a function of $\pt^\Pgg$ measured in data over the corresponding GamJet NLO theoretical prediction (obtained with the CJ15 PDFs) in different bins of $\absetajet$ for $\abs{\eta^\Pgg}<0.8$.  Error bars on the data are statistical uncertainties, and blue bands represent the systematic uncertainties.}
\label{fig:Ratio0}
\end{figure*}

\begin{figure*}[htb!]
\centering
\includegraphics[width=\cmsFigWidthiii]{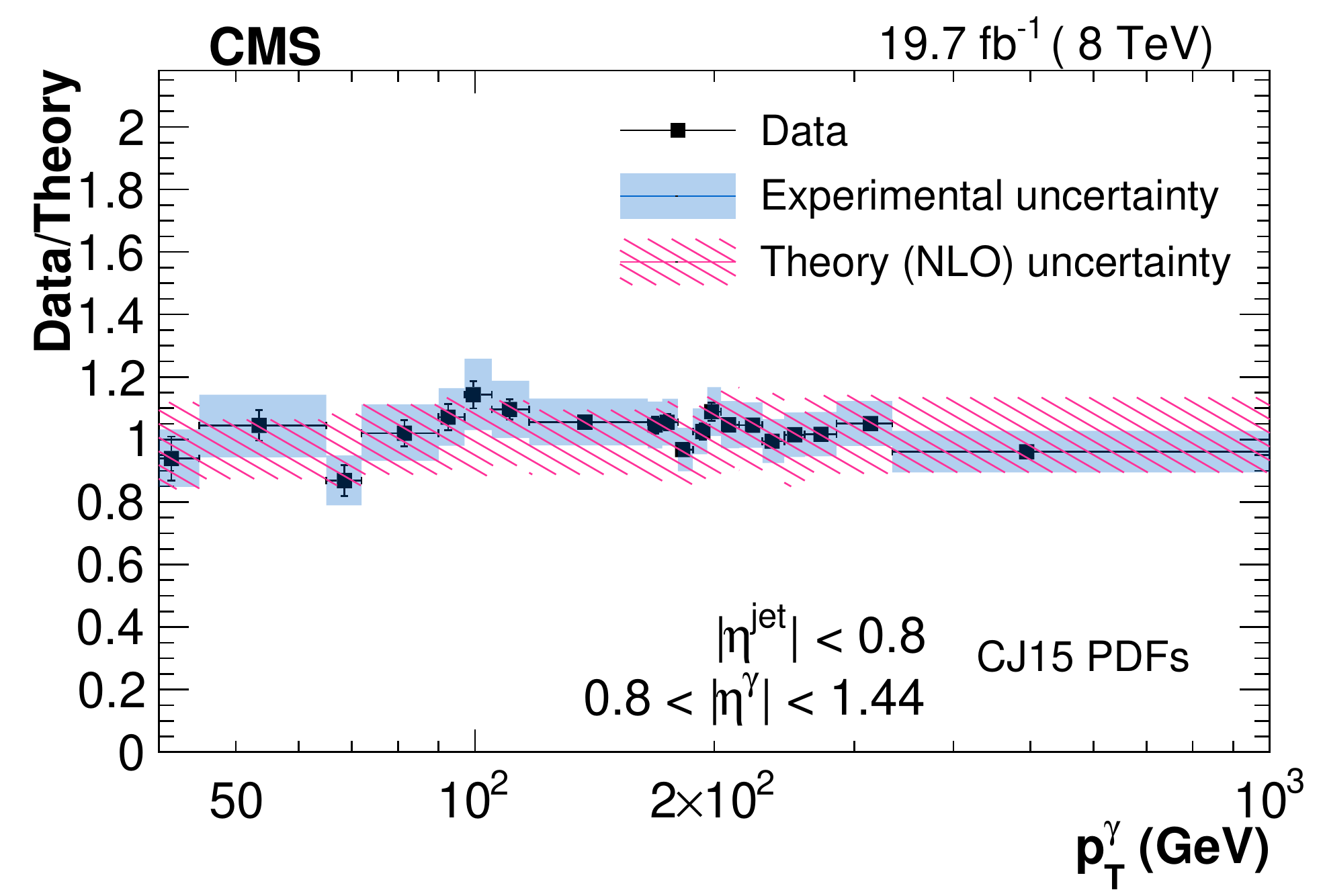}
\includegraphics[width=\cmsFigWidthiii]{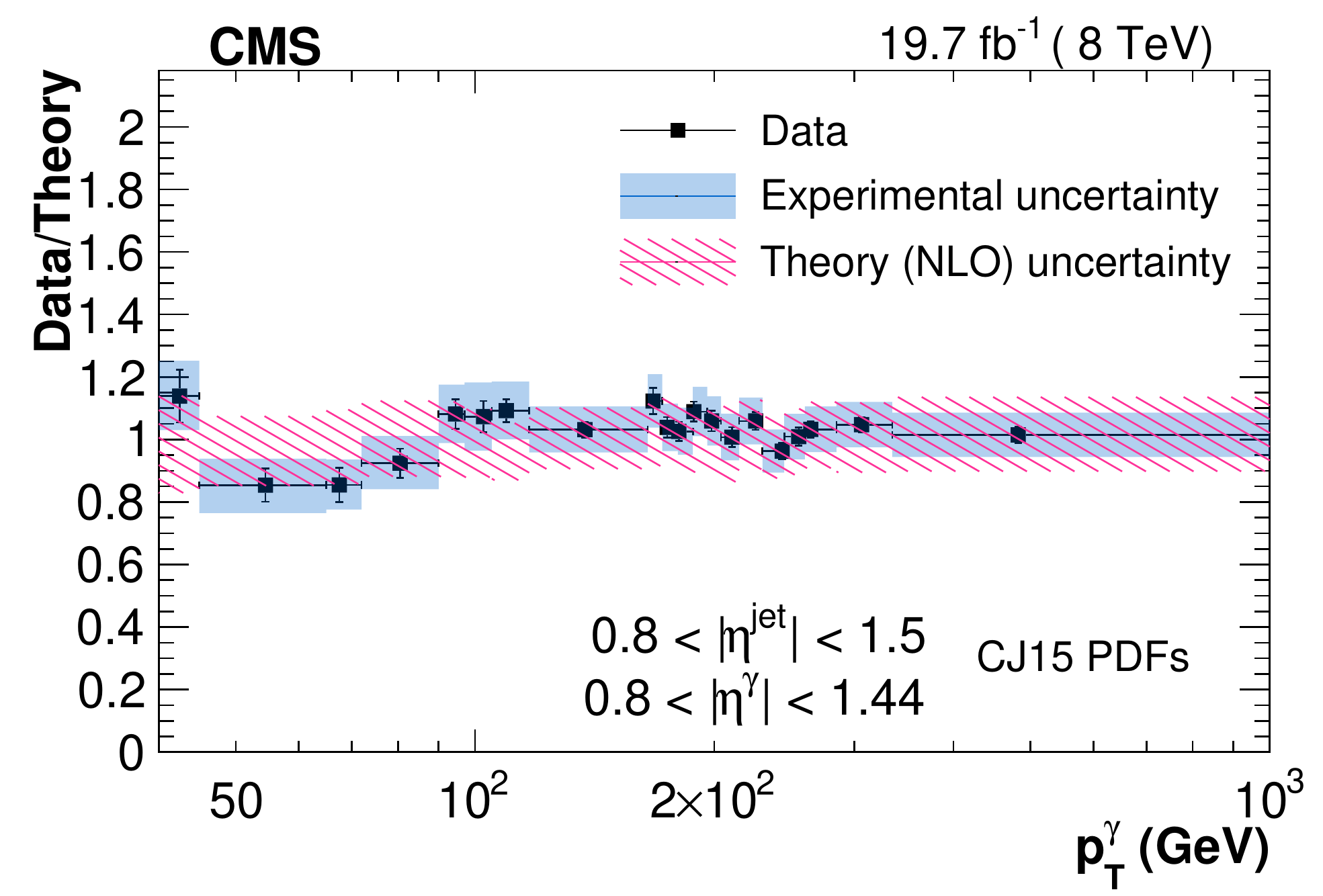}
\includegraphics[width=\cmsFigWidthiii]{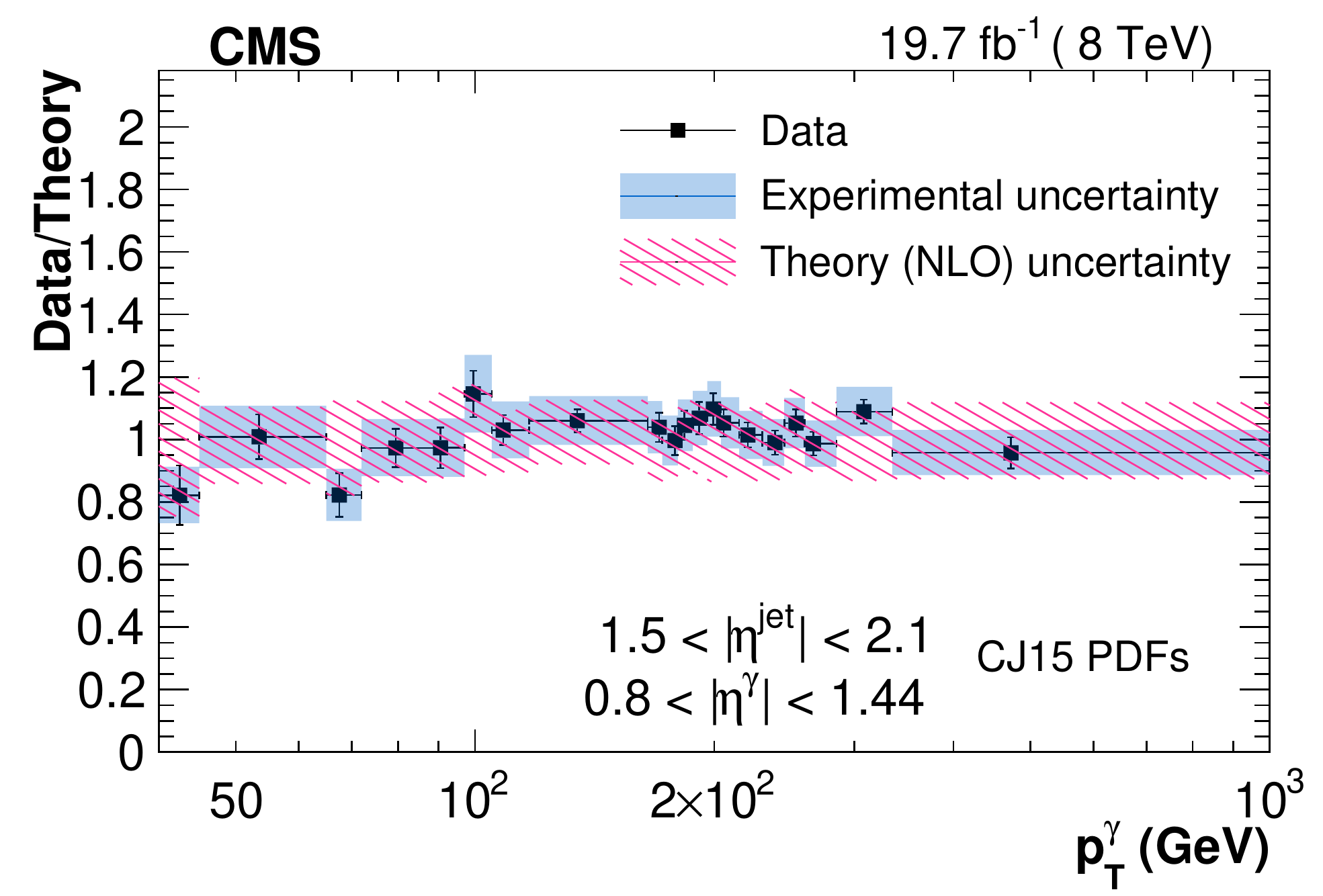}
\includegraphics[width=\cmsFigWidthiii]{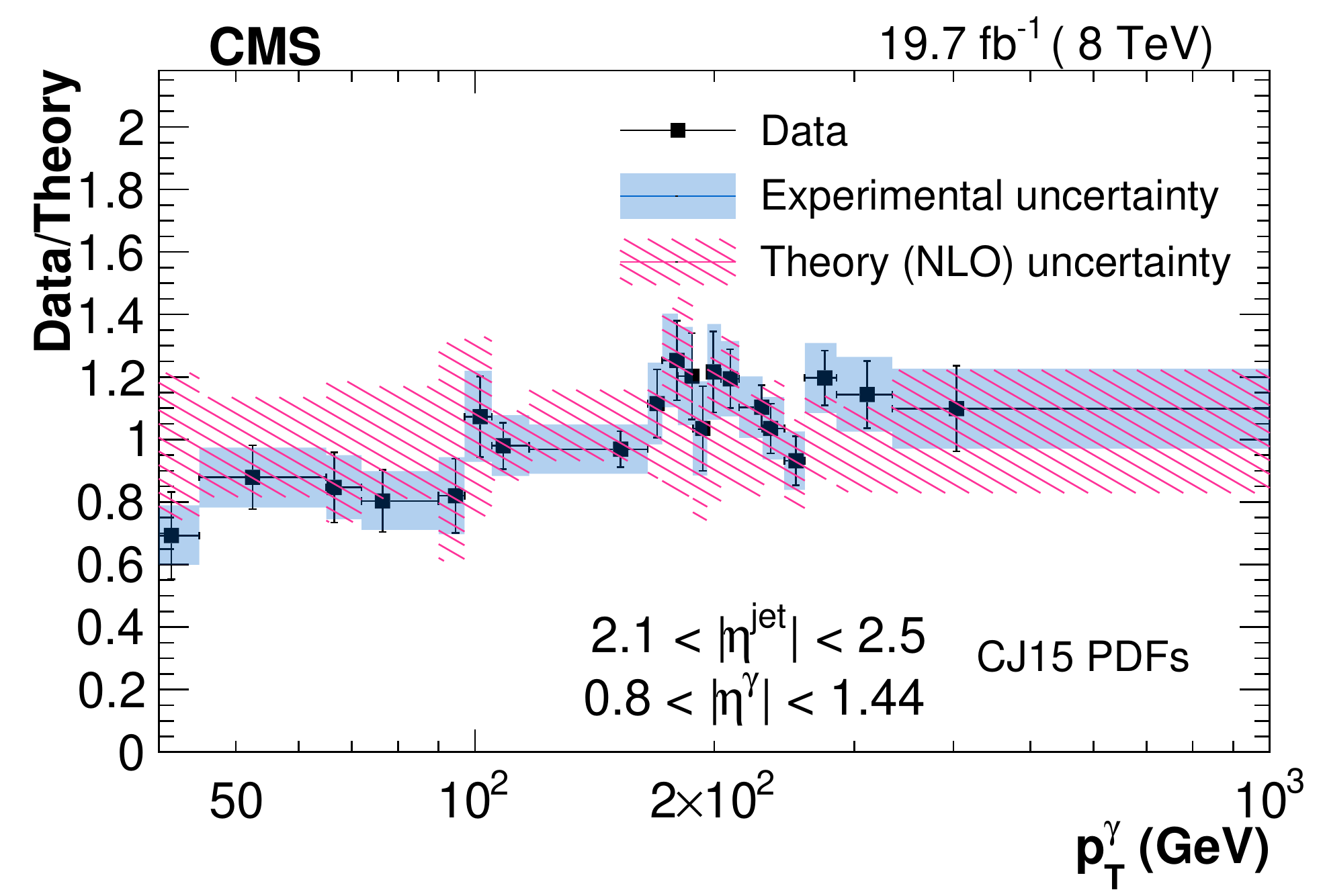}
\caption{Ratio of triple-differential cross sections as a function of $\pt^\Pgg$ measured in data over the corresponding GamJet NLO theoretical prediction (obtained with the CJ15 PDFs) in different bins of $\absetajet$ for $0.80<\abs{\eta^\Pgg}<1.44$.  Error bars on the data are statistical uncertainties, and blue bands represent the systematic uncertainties.}
\label{fig:Ratio1}
\end{figure*}

\begin{figure*}[htb!]
\centering
\includegraphics[width=\cmsFigWidthiii]{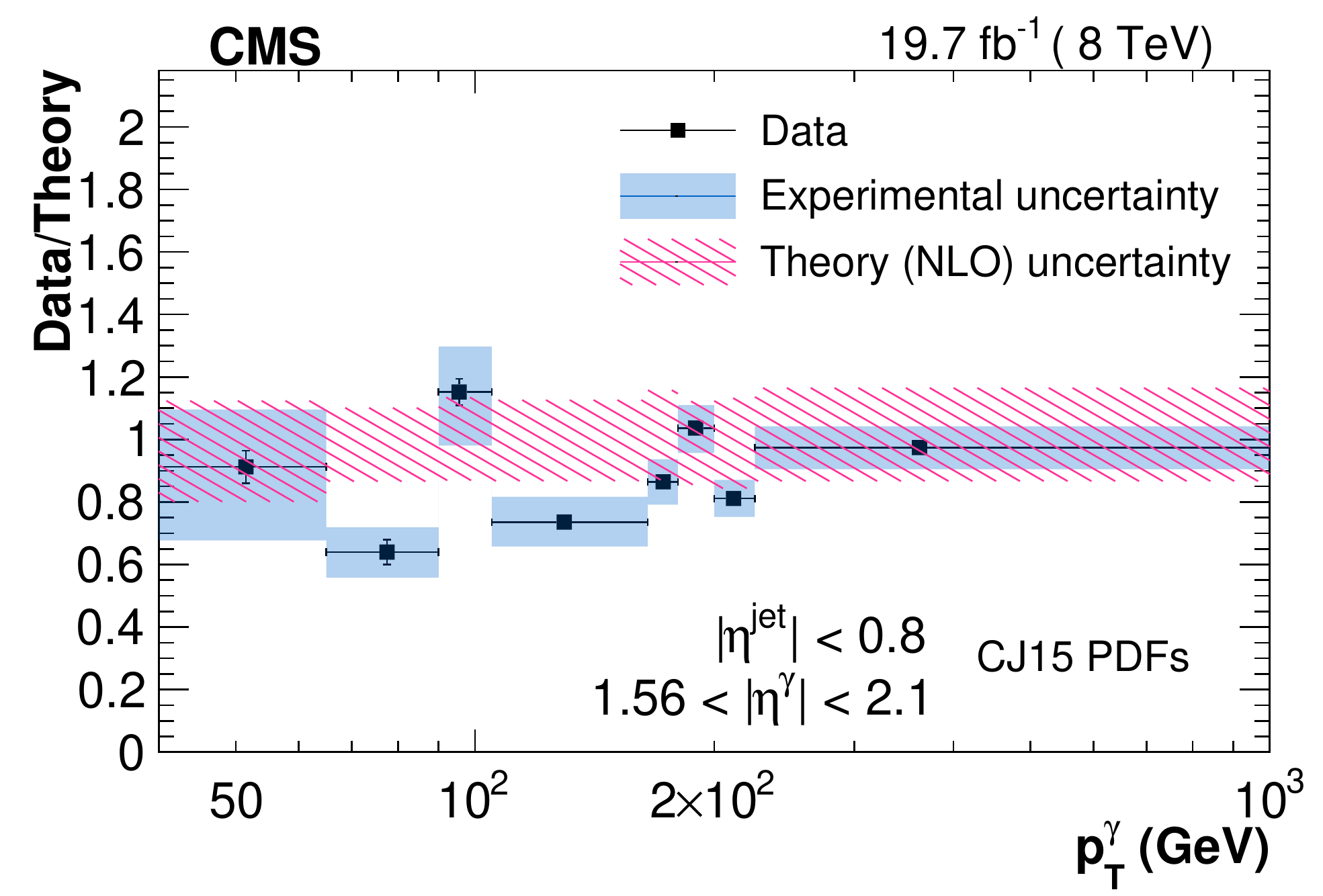}
\includegraphics[width=\cmsFigWidthiii]{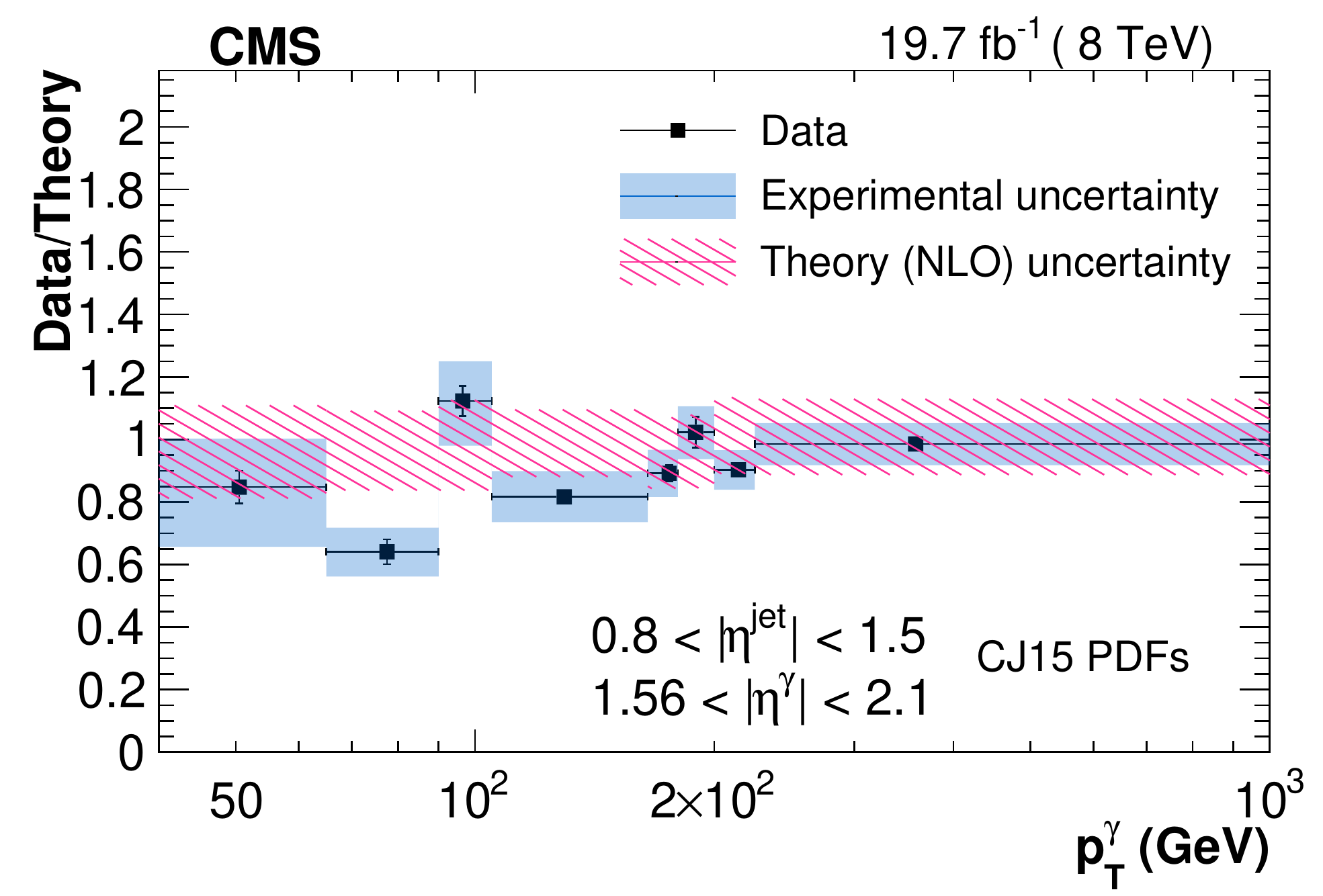}
\includegraphics[width=\cmsFigWidthiii]{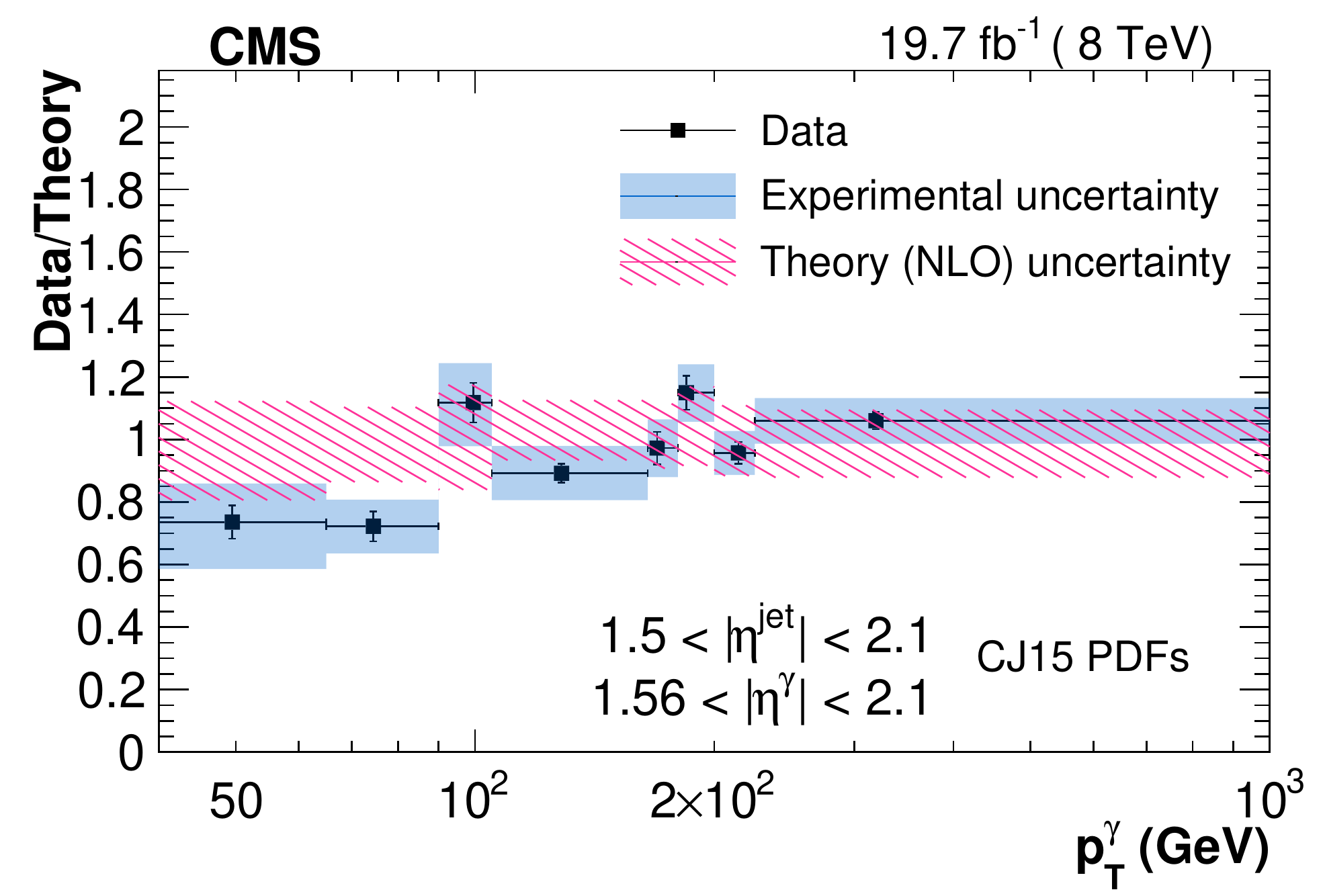}
\includegraphics[width=\cmsFigWidthiii]{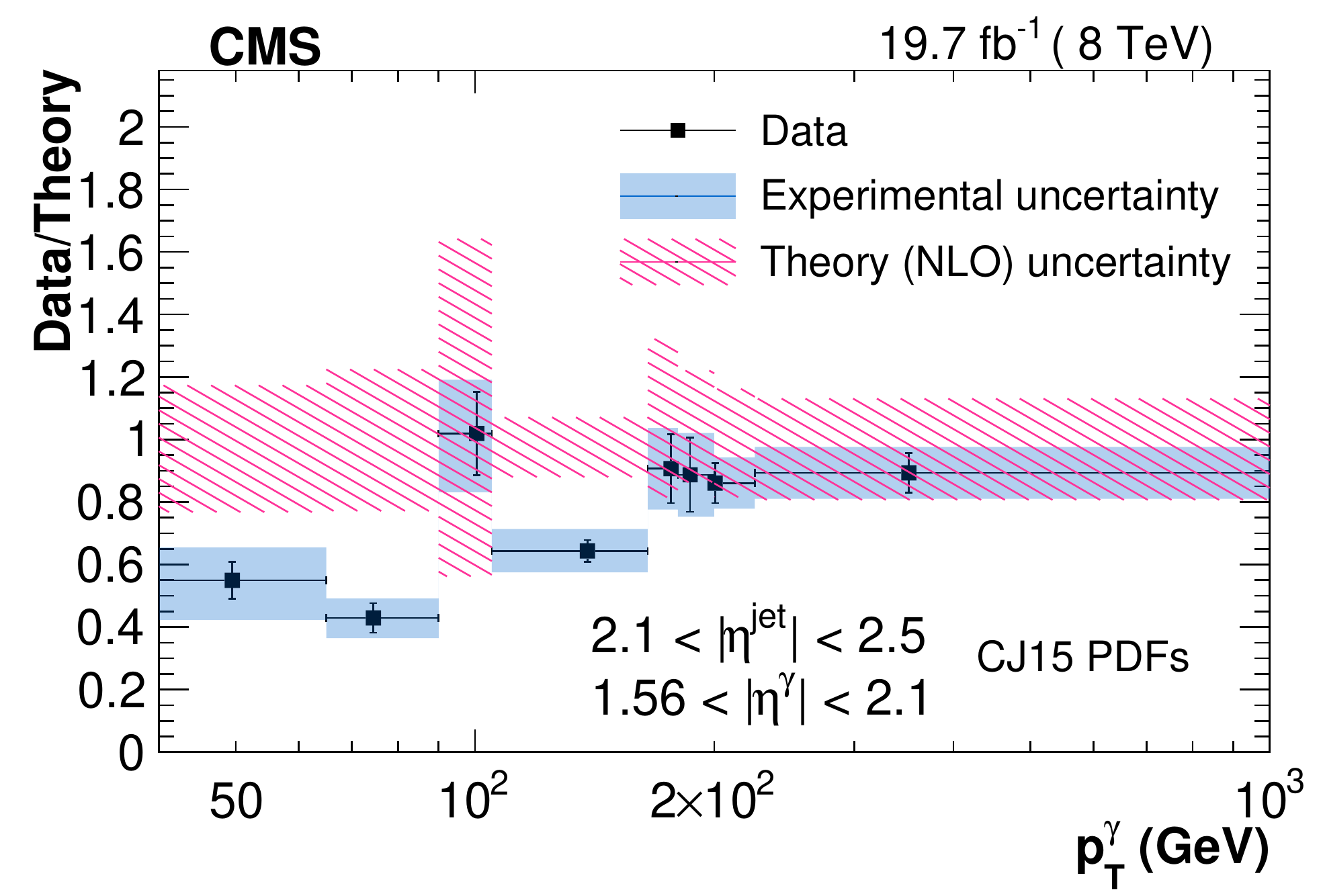}
\caption{Ratio of triple-differential cross sections as a function of $\pt^\Pgg$ measured in data over the corresponding GamJet NLO theoretical prediction (obtained with the CJ15 PDFs) in different bins of $\absetajet$ for $1.56<\abs{\eta^\Pgg}<2.10$.  Error bars on the data are statistical uncertainties, and blue bands represent the systematic uncertainties.}
\label{fig:Ratio2}
\end{figure*}

\begin{figure*}[htb!]
\centering
\includegraphics[width=\cmsFigWidthiii]{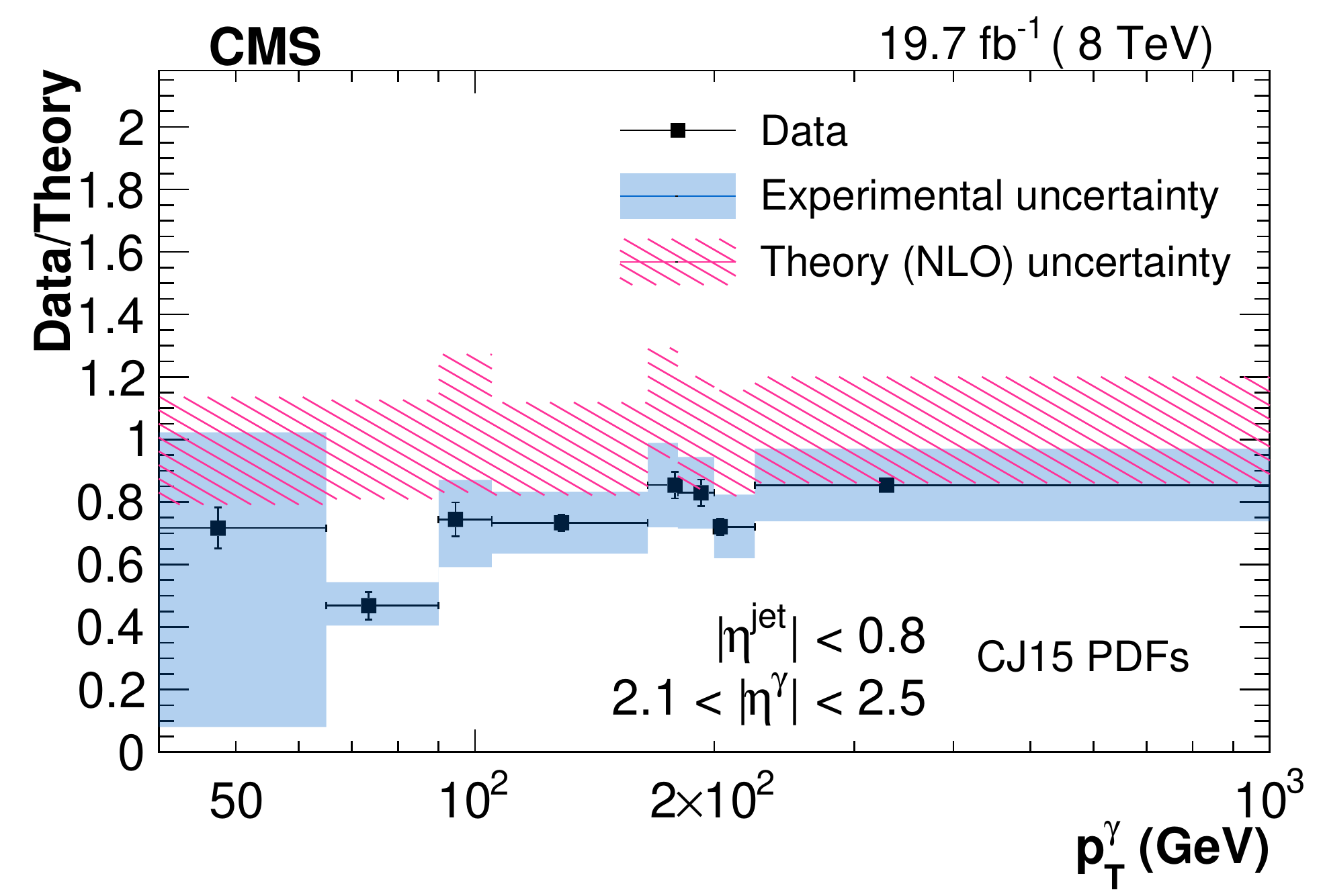}
\includegraphics[width=\cmsFigWidthiii]{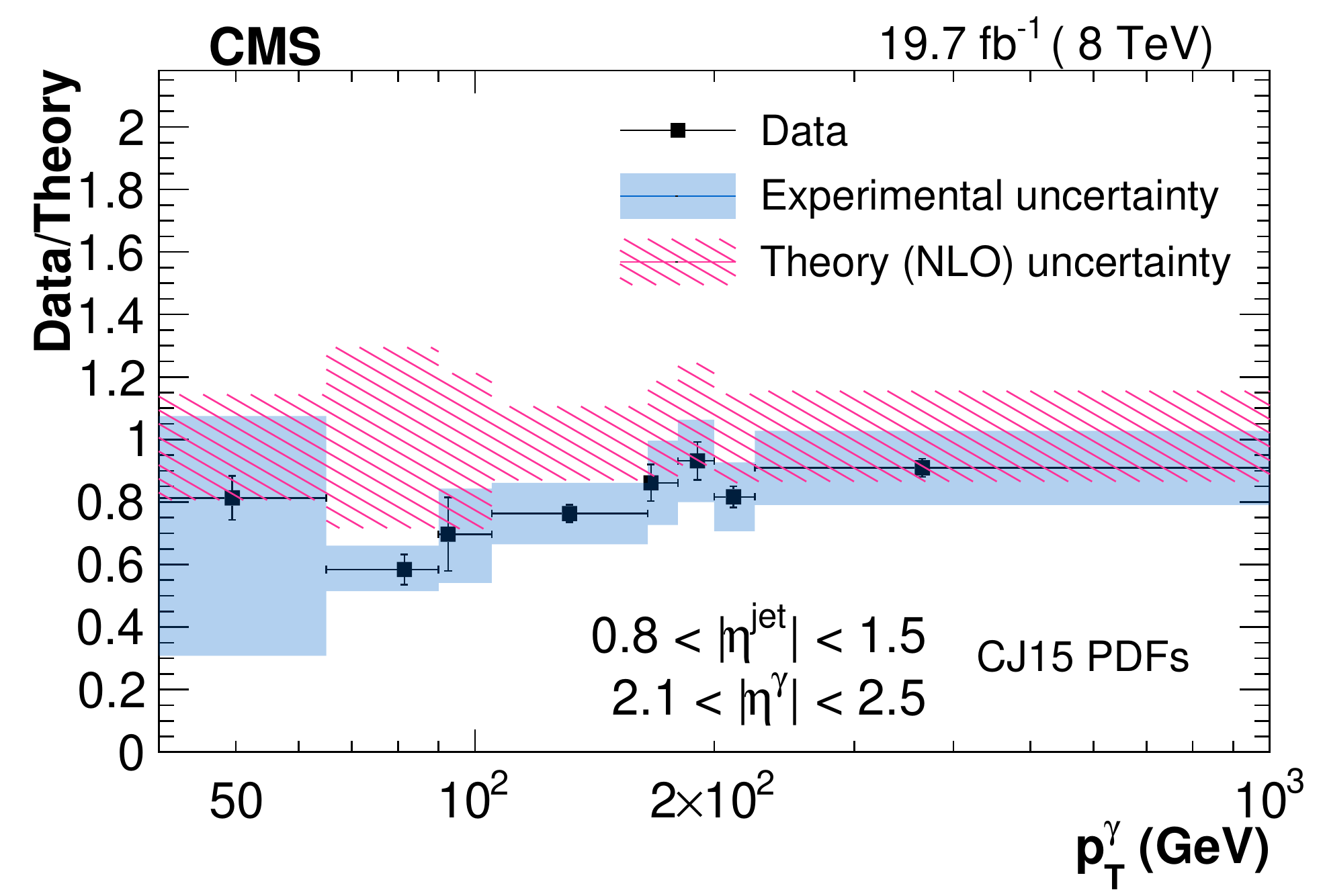}
\includegraphics[width=\cmsFigWidthiii]{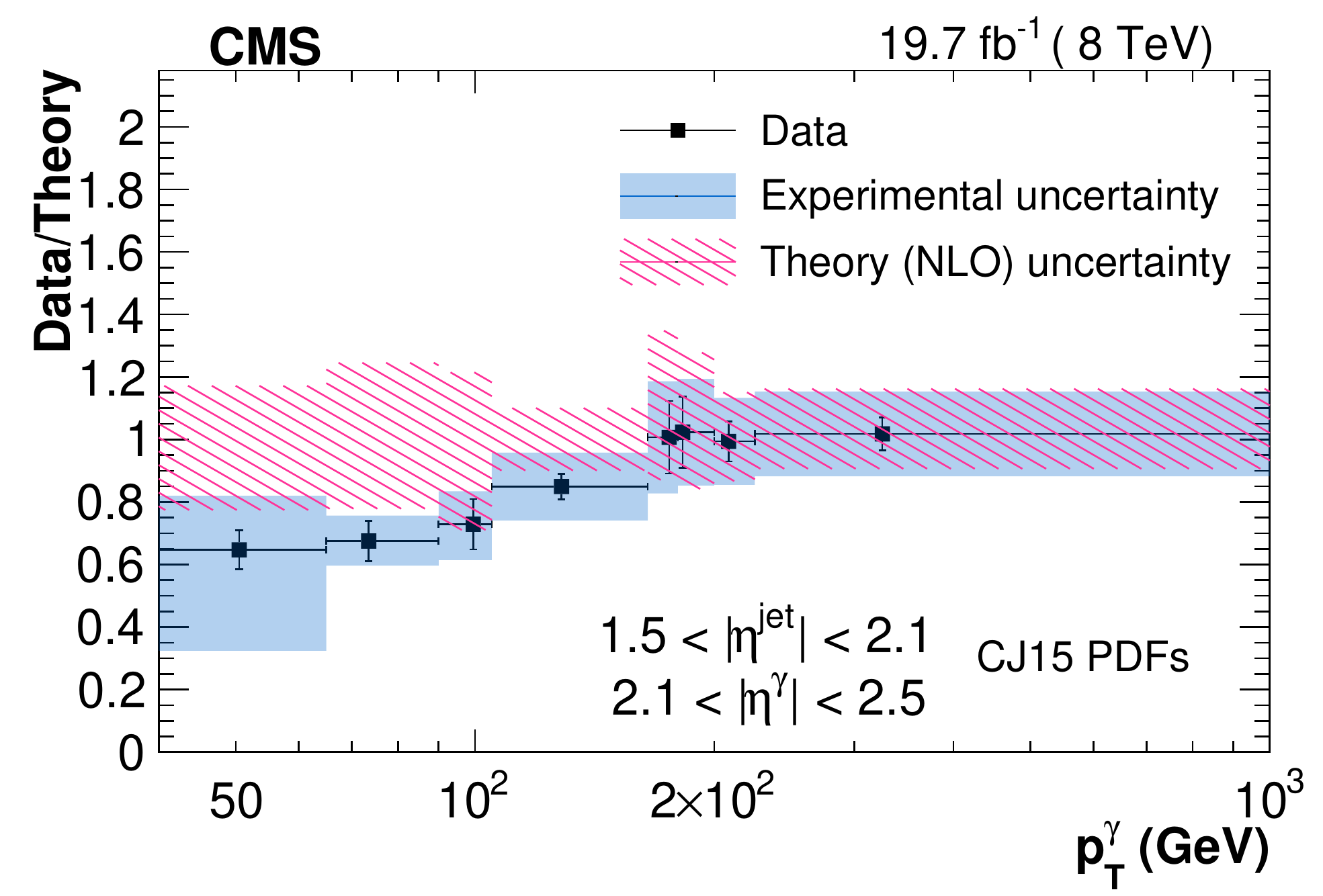}
\includegraphics[width=\cmsFigWidthiii]{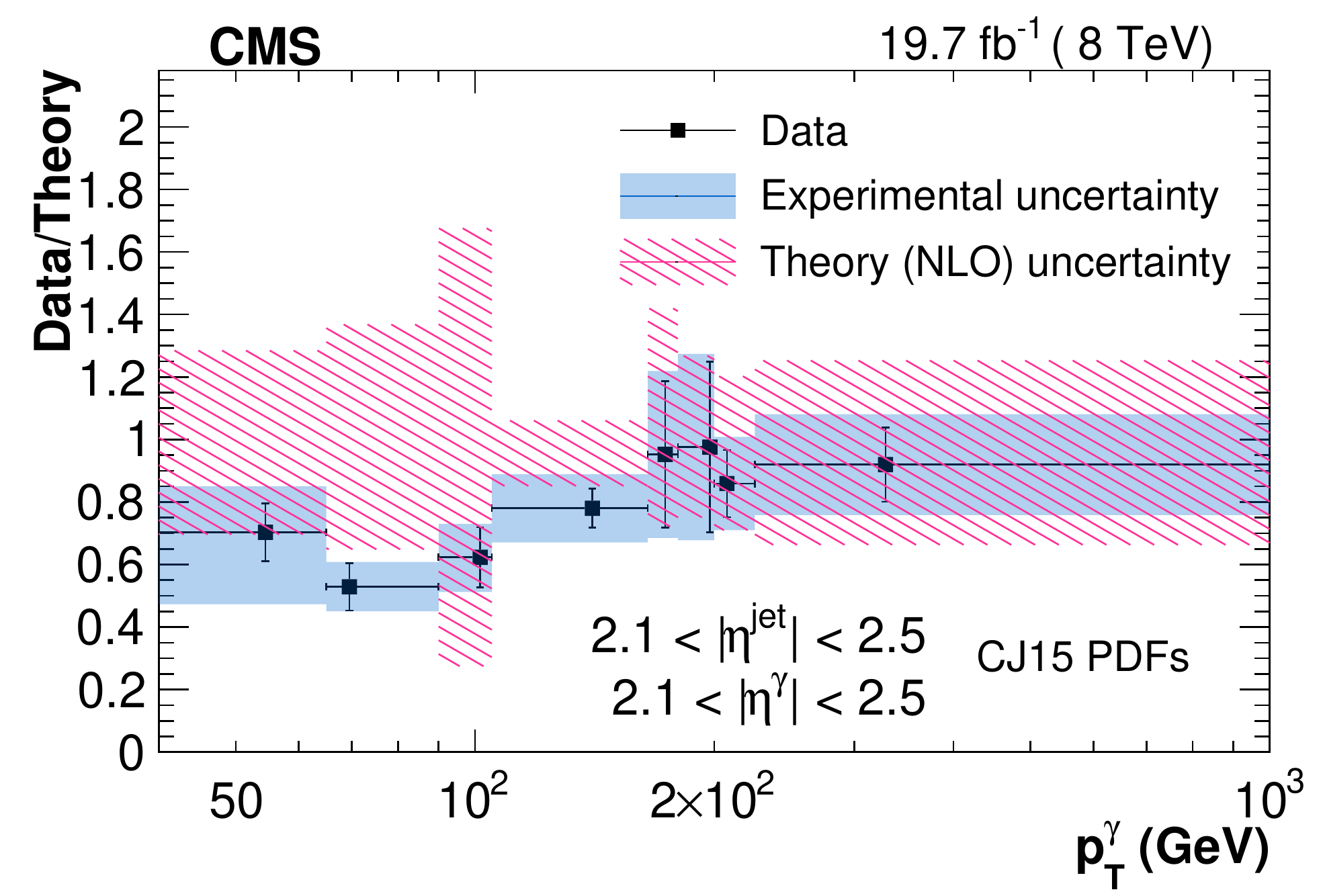}
\caption{Ratio of triple-differential cross sections as a function of $\pt^\Pgg$ measured in data over the corresponding GamJet NLO theoretical prediction (obtained with the CJ15 PDFs) in different bins of $\absetajet$ for $2.1<\abs{\eta^\Pgg}<2.5$.  Error bars on the data are statistical uncertainties, and blue bands represent the systematic uncertainties.}
\label{fig:Ratio3}
\end{figure*}

The measured triple-differential cross sections are shown in
Figs.~\ref{fig:FinalXSecCen} and~\ref{fig:FinalXSecFwd}.
A summary of the uncertainty in the measured cross sections from different sources
is reported in Table~\ref{tab:crosssecun}.
Comparison between data and theory, along with the respective uncertainties, are provided in Figs.~\ref{fig:Ratio0}--\ref{fig:Ratio3}.
The measurements are in good agreement with the NLO QCD predictions from GamJet
except in the regions of low $\ptgamma$ for endcap photons, where differences of up
to 60\% are observed between central values of the data and theoretical predictions.
\begin{table}[htb]
 \centering
\topcaption {Summary of the uncertainties in the measured cross section values for photons in the barrel (endcap) region.}
  \begin{tabular}{ l  c  c}
   Sources              &	Barrel photons		&	Endcap photons\\\hline
   Statistical	        &	1--20	\%	&	1--10	\%	\\
   Purity	        &	1--9	\%	&	3--66	\%	\\
   Efficiency	        &	1--9	\%	&	5--11	\%	\\
   Luminosity	        &	3	\%	&	3	\%	\\
   Unfolding	        &	0--5	\%	&	0--1	\%	\\
   Total systematic	&	4--12	\%	&	6--66	\%	\\
  \end{tabular}
\label{tab:crosssecun}
\end{table}

\section{Summary}\label{summary}
Measurements of the triple-differential inclusive isolated-photon+jet
cross section were performed as a function of
photon transverse momentum ($\ptgamma$), photon pseudorapidity
($\etagamma$), and jet pseudorapidity ($\etajet$).
The measurements were carried out in \Pp\Pp collisions at $\sqrt{s} = 8\TeV$
using 19.7\fbinv of data collected by the CMS detector covering a
kinematic range of $\absetagamma<1.44$ and $1.57<\absetagamma<2.50$, $\absetajet<2.5$,
$40< \ptgamma<1000$\GeV, and jet transverse momentum, $\ptjet$, $>$25\GeV.
The photon purity was estimated using a combination of
templates from data and simulation, based on a multivariate technique.
The measured cross sections are in good agreement with the
next-to-leading order perturbative quantum chromodynamics (pQCD) prediction, and the
experimental uncertainties are comparable or smaller
than the theoretical ones.  These measured cross sections, in different combinations
of photon and jet pseudorapidities, probe pQCD over a wide range
of parton momentum fractions.  Inclusion of such gluon-sensitive data
into the global parton distribution function (PDF) fit analyses has the
potential to constrain the gluon PDFs,
particularly in the regions where the measured uncertainties
are smaller than the uncertainty bands of theoretical predictions.

\clearpage

\begin{acknowledgments}
We congratulate our colleagues in the CERN accelerator departments for the excellent performance of the LHC and thank the technical and administrative staffs at CERN and at other CMS institutes for their contributions to the success of the CMS effort. In addition, we gratefully acknowledge the computing centers and personnel of the Worldwide LHC Computing Grid for delivering so effectively the computing infrastructure essential to our analyses. Finally, we acknowledge the enduring support for the construction and operation of the LHC and the CMS detector provided by the following funding agencies: BMBWF and FWF (Austria); FNRS and FWO (Belgium); CNPq, CAPES, FAPERJ, FAPERGS, and FAPESP (Brazil); MES (Bulgaria); CERN; CAS, MoST, and NSFC (China); COLCIENCIAS (Colombia); MSES and CSF (Croatia); RPF (Cyprus); SENESCYT (Ecuador); MoER, ERC IUT, PUT and ERDF (Estonia); Academy of Finland, MEC, and HIP (Finland); CEA and CNRS/IN2P3 (France); BMBF, DFG, and HGF (Germany); GSRT (Greece); NKFIA (Hungary); DAE and DST (India); IPM (Iran); SFI (Ireland); INFN (Italy); MSIP and NRF (Republic of Korea); MES (Latvia); LAS (Lithuania); MOE and UM (Malaysia); BUAP, CINVESTAV, CONACYT, LNS, SEP, and UASLP-FAI (Mexico); MOS (Montenegro); MBIE (New Zealand); PAEC (Pakistan); MSHE and NSC (Poland); FCT (Portugal); JINR (Dubna); MON, RosAtom, RAS, RFBR, and NRC KI (Russia); MESTD (Serbia); SEIDI, CPAN, PCTI, and FEDER (Spain); MOSTR (Sri Lanka); Swiss Funding Agencies (Switzerland); MST (Taipei); ThEPCenter, IPST, STAR, and NSTDA (Thailand); TUBITAK and TAEK (Turkey); NASU and SFFR (Ukraine); STFC (United Kingdom); DOE and NSF (USA).

\hyphenation{Rachada-pisek} Individuals have received support from the Marie-Curie program and the European Research Council and Horizon 2020 Grant, contract Nos.\ 675440, 752730, and 765710 (European Union); the Leventis Foundation; the A.P.\ Sloan Foundation; the Alexander von Humboldt Foundation; the Belgian Federal Science Policy Office; the Fonds pour la Formation \`a la Recherche dans l'Industrie et dans l'Agriculture (FRIA-Belgium); the Agentschap voor Innovatie door Wetenschap en Technologie (IWT-Belgium); the F.R.S.-FNRS and FWO (Belgium) under the ``Excellence of Science -- EOS" -- be.h project n.\ 30820817; the Beijing Municipal Science \& Technology Commission, No. Z181100004218003; the Ministry of Education, Youth and Sports (MEYS) of the Czech Republic; the Lend\"ulet (``Momentum") Program and the J\'anos Bolyai Research Scholarship of the Hungarian Academy of Sciences, the New National Excellence Program \'UNKP, the NKFIA research grants 123842, 123959, 124845, 124850, 125105, 128713, 128786, and 129058 (Hungary); the Council of Science and Industrial Research, India; the HOMING PLUS program of the Foundation for Polish Science, cofinanced from European Union, Regional Development Fund, the Mobility Plus program of the Ministry of Science and Higher Education, the National Science Center (Poland), contracts Harmonia 2014/14/M/ST2/00428, Opus 2014/13/B/ST2/02543, 2014/15/B/ST2/03998, and 2015/19/B/ST2/02861, Sonata-bis 2012/07/E/ST2/01406; the National Priorities Research Program by Qatar National Research Fund; the Ministry of Science and Education, grant no. 3.2989.2017 (Russia); the Programa Estatal de Fomento de la Investigaci{\'o}n Cient{\'i}fica y T{\'e}cnica de Excelencia Mar\'{\i}a de Maeztu, grant MDM-2015-0509 and the Programa Severo Ochoa del Principado de Asturias; the Thalis and Aristeia programs cofinanced by EU-ESF and the Greek NSRF; the Rachadapisek Sompot Fund for Postdoctoral Fellowship, Chulalongkorn University and the Chulalongkorn Academic into Its 2nd Century Project Advancement Project (Thailand); the Welch Foundation, contract C-1845; and the Weston Havens Foundation (USA). \end{acknowledgments}

\bibliography{auto_generated}

\providecommand{\href}[2]{#2}\begingroup\raggedright\begin{thebibliography}{10}%
\makeatletter
\providecommand{\hrefCMSnoop }[0]{\@secondoftwo}%
\makeatother
\providecommand{\doi}{\texttt{doi:}\begingroup \urlstyle{tt}\Url}

\bibitem{Campbell:2016lzl}
\hrefCMSnoop {}{J.~M. Campbell, R.~K. Ellis, and C.~Williams, ``{Direct photon
  production at next-to-next-to-leading order}'',} \textit{ Phys. Rev. Lett.}
  \textbf{ 118} (2017) 222001,
  \href{http://dx.doi.org/10.1103/PhysRevLett.118.222001}{\doi{10.1103/PhysRevLett.118.222001}},
\href{http://www.arXiv.org/abs/1612.04333}{\texttt{arXiv:1612.04333}}.

\bibitem{Chen:2019zmr}
X.~Chen\hrefCMSnoop {}{ {et~al.}, ``{Isolated photon and photon+jet production
  at NNLO QCD accuracy}'',} \textit{ Submitted to: JHEP} (2019)
\href{http://www.arXiv.org/abs/1904.01044}{\texttt{arXiv:1904.01044}}.

\bibitem{dEnterria:2012kvo}
\hrefCMSnoop {}{D.~d'Enterria and J.~Rojo, ``{Quantitative constraints on the
  gluon distribution function in the proton from collider isolated-photon
  data}'',} \textit{ Nucl. Phys. B} \textbf{ {860}} (2012) 311,
  \href{http://dx.doi.org/10.1016/j.nuclphysb.2012.03.003}{\doi{10.1016/j.nuclphysb.2012.03.003}},
\href{http://www.arXiv.org/abs/1202.1762}{\texttt{arXiv:1202.1762}}.

\bibitem{Carminati:2012mm}
L.~Carminati\hrefCMSnoop {}{ {et~al.}, ``{Sensitivity of the LHC
  isolated-gamma+jet data to the parton distribution functions of the
  proton}'',} \textit{ EPL} \textbf{ 101} (2013) 61002,
  \href{http://dx.doi.org/10.1209/0295-5075/101/61002}{\doi{10.1209/0295-5075/101/61002}},
\href{http://www.arXiv.org/abs/1212.5511}{\texttt{arXiv:1212.5511}}.

\bibitem{Campbell:2018wfu}
\hrefCMSnoop {}{J.~M. Campbell, J.~Rojo, E.~Slade, and C.~Williams, ``{Direct
  photon production and PDF fits reloaded}'',} \textit{ Eur. Phys. J. C}
  \textbf{ {78}} (2018) 470,
  \href{http://dx.doi.org/10.1140/epjc/s10052-018-5944-4}{\doi{10.1140/epjc/s10052-018-5944-4}},
\href{http://www.arXiv.org/abs/1802.03021}{\texttt{arXiv:1802.03021}}.

\bibitem{Aad:2016xcr}
\hrefCMSnoop {}{{ATLAS Collaboration}, ``{Measurement of the inclusive isolated
  prompt photon cross section in pp collisions at $ \sqrt{s}=8 $ TeV with the
  ATLAS detector}'',} \textit{ JHEP} \textbf{ 08} (2016) 005,
  \href{http://dx.doi.org/10.1007/JHEP08(2016)005}{\doi{10.1007/JHEP08(2016)005}},
\href{http://www.arXiv.org/abs/1605.03495}{\texttt{arXiv:1605.03495}}.

\bibitem{Aad:2013gaa}
\hrefCMSnoop {}{{ATLAS Collaboration}, ``{Dynamics of isolated-photon plus jet
  production in pp collisions at $\sqrt{s}=7\TeV$ with the ATLAS detector}'',}
  \textit{ Nucl. Phys. B} \textbf{ 875} (2013) 483,
  \href{http://dx.doi.org/10.1016/j.nuclphysb.2013.07.025}{\doi{10.1016/j.nuclphysb.2013.07.025}},
\href{http://www.arXiv.org/abs/1307.6795}{\texttt{arXiv:1307.6795}}.

\bibitem{Chatrchyan:2013mwa}
\hrefCMSnoop {}{{CMS Collaboration}, ``{Measurement of the triple-differential
  cross section for photon+jets production in proton-proton collisions at
  $\sqrt{s} = 7 \TeV$}'',} \textit{ JHEP} \textbf{ 06} (2014) 009,
  \href{http://dx.doi.org/10.1007/JHEP06(2014)009}{\doi{10.1007/JHEP06(2014)009}},
\href{http://www.arXiv.org/abs/1311.6141}{\texttt{arXiv:1311.6141}}.

\bibitem{D0:2013lra}
\hrefCMSnoop {}{{D0} Collaboration, ``{Measurement of the differential cross
  section of photon plus jet production in $\mathrm{p\bar{p}}$ collisions at
  $\sqrt{s} = 1.96\TeV$}'',} \textit{ Phys. Rev. D} \textbf{ 88} (2013) 072008,
  \href{http://dx.doi.org/10.1103/PhysRevD.88.072008}{\doi{10.1103/PhysRevD.88.072008}},
\href{http://www.arXiv.org/abs/1308.2708}{\texttt{arXiv:1308.2708}}.

\bibitem{ATLAS:2012ar}
\hrefCMSnoop {}{{ATLAS Collaboration}, ``{Measurement of the production cross
  section of an isolated photon associated with jets in proton-proton
  collisions at $\sqrt{s}=7$ TeV with the ATLAS detector}'',} \textit{ Phys.
  Rev. D} \textbf{ 85} (2012) 092014,
  \href{http://dx.doi.org/10.1103/PhysRevD.85.092014}{\doi{10.1103/PhysRevD.85.092014}},
\href{http://www.arXiv.org/abs/1203.3161}{\texttt{arXiv:1203.3161}}.

\bibitem{Abazov:2008er}
\hrefCMSnoop {}{{D0} Collaboration, ``{Measurement of the differential cross
  section for the production of an isolated photon with associated jet in
  $\mathrm{p\bar{p}}$ collisions at $\sqrt{s} = 1.96\TeV$}'',} \textit{ Phys.
  Lett. B} \textbf{ 666} (2008) 435,
  \href{http://dx.doi.org/10.1016/j.physletb.2008.06.076}{\doi{10.1016/j.physletb.2008.06.076}},
\href{http://www.arXiv.org/abs/0804.1107}{\texttt{arXiv:0804.1107}}.

\bibitem{Chatrchyan:2008zzk}
\hrefCMSnoop {}{{CMS Collaboration}, ``The {CMS} experiment at the {CERN}
  {LHC}'',} \textit{ JINST} \textbf{ 3} (2008) S08004,
\href{http://dx.doi.org/10.1088/1748-0221/3/08/S08004}{\doi{10.1088/1748-0221/3/08/S08004}}.

\bibitem{1748-0221-12-10-P10003}
\hrefCMSnoop {}{{CMS Collaboration}, ``{Particle-flow reconstruction and global
  event description with the CMS detector}'',} \textit{ JINST} \textbf{ 12}
  (2017) P10003,
  \href{http://dx.doi.org/10.1088/1748-0221/12/10/p10003}{\doi{10.1088/1748-0221/12/10/p10003}},
  \href{http://www.arXiv.org/abs/1706.04965}{\texttt{arXiv:1706.04965}}.

\bibitem{Cacciari:2008gp}
\hrefCMSnoop {}{M.~Cacciari, G.~P. Salam, and G.~Soyez, ``{The
  anti-$k_\mathrm{T}$ jet clustering algorithm}'',} \textit{ JHEP} \textbf{ 04}
  (2008) 063,
  \href{http://dx.doi.org/10.1088/1126-6708/2008/04/063}{\doi{10.1088/1126-6708/2008/04/063}},
  \href{http://www.arXiv.org/abs/0802.1189}{\texttt{arXiv:0802.1189}}.

\bibitem{JESPaper}
\hrefCMSnoop {}{{CMS Collaboration}, ``{Jet energy scale and resolution in the
  CMS experiment in pp collisions at 8 TeV}'',} \textit{ JINST} \textbf{ 12}
  (2017) P02014,
  \href{http://dx.doi.org/10.1088/1748-0221/12/02/P02014}{\doi{10.1088/1748-0221/12/02/P02014}},
  \href{http://www.arXiv.org/abs/1607.03663}{\texttt{arXiv:1607.03663}}.

\bibitem{CMS-PAS-JME-16-003}
\href {https://cds.cern.ch/record/2256875}{{CMS Collaboration}, ``Jet
  algorithms performance in 13 {TeV} data'',} CMS Physics Analysis Summary
  CMS-PAS-JME-16-003, {CERN}, 2017.

\bibitem{CMS:EGM-14-001}
\hrefCMSnoop {}{{CMS Collaboration}, ``{Performance of photon reconstruction
  and identification with the CMS detector in proton-proton collisions at
  $\sqrt{s} = 8$ TeV}'',} \textit{ JINST} \textbf{ 10} (2015) P08010,
  \href{http://dx.doi.org/10.1088/1748-0221/10/08/P08010}{\doi{10.1088/1748-0221/10/08/P08010}},
\href{http://www.arXiv.org/abs/1502.02702}{\texttt{arXiv:1502.02702}}.

\bibitem{Chatrchyan:2012xi}
\hrefCMSnoop {}{{CMS Collaboration}, ``{Performance of CMS muon reconstruction
  in pp collision events at $\sqrt{s}=7$ TeV}'',} \textit{ JINST} \textbf{ 7}
  (2012) P10002,
  \href{http://dx.doi.org/10.1088/1748-0221/7/10/P10002}{\doi{10.1088/1748-0221/7/10/P10002}},
\href{http://www.arXiv.org/abs/1206.4071}{\texttt{arXiv:1206.4071}}.

\bibitem{1748-0221-12-01-P01020}
\hrefCMSnoop {}{{CMS Collaboration}, ``{The CMS trigger system}'',} \textit{
  JINST} \textbf{ 12} (2017) P01020,
  \href{http://dx.doi.org/10.1088/1748-0221/12/01/P01020}{\doi{10.1088/1748-0221/12/01/P01020}},
  \href{http://www.arXiv.org/abs/1609.02366}{\texttt{arXiv:1609.02366}}.

\bibitem{1126-6708-2006-05-026}
\hrefCMSnoop {}{T.~Sj{\"o}strand, S.~Mrenna, and P.~Skands, ``{{\sc pythia} 6.4
  physics and manual}'',} \textit{ JHEP} \textbf{ 05} (2006) 026,
  \href{http://dx.doi.org/10.1088/1126-6708/2006/05/026}{\doi{10.1088/1126-6708/2006/05/026}},
\href{http://www.arXiv.org/abs/hep-ph/0603175}{\texttt{arXiv:hep-ph/0603175}}.

\bibitem{Agostinelli:2002hh}
\hrefCMSnoop {}{{GEANT4} Collaboration, ``{\GEANTfour}---a simulation
  toolkit'',} \textit{ Nucl. Instrum. Meth. A} \textbf{ 506} (2003) 250,
\href{http://dx.doi.org/10.1016/S0168-9002(03)01368-8}{\doi{10.1016/S0168-9002(03)01368-8}}.

\bibitem{1126-6708-2003-02-027}
\hrefCMSnoop {}{F.~Maltoni and T.~Stelzer, ``{MadEvent: automatic event
  generation with MadGraph}'',} \textit{ JHEP} \textbf{ 02} (2003) 027,
  \href{http://dx.doi.org/10.1088/1126-6708/2003/02/027}{\doi{10.1088/1126-6708/2003/02/027}},
  \href{http://www.arXiv.org/abs/hep-ph/0208156}{\texttt{arXiv:hep-ph/0208156}}.

\bibitem{Pumplin:2002vw}
J.~Pumplin\hrefCMSnoop {}{ {et~al.}, ``{New generation of parton distributions
  with uncertainties from global QCD analysis}'',} \textit{ JHEP} \textbf{ 07}
  (2002) 012,
  \href{http://dx.doi.org/10.1088/1126-6708/2002/07/012}{\doi{10.1088/1126-6708/2002/07/012}},
\href{http://www.arXiv.org/abs/hep-ph/0201195}{\texttt{arXiv:hep-ph/0201195}}.

\bibitem{Chatrchyan2013}
\hrefCMSnoop {}{{CMS Collaboration}, ``{Study of the underlying event at
  forward rapidity in pp collisions at $\sqrt{s}$ = 0.9, 2.76, and 7 TeV}'',}
  \textit{ JHEP} \textbf{ 04} (2013) 072,
  \href{http://dx.doi.org/10.1007/JHEP04(2013)072}{\doi{10.1007/JHEP04(2013)072}}.

\bibitem{ROE2005577}
B.~P. Roe\hrefCMSnoop {}{ {et~al.}, ``{Boosted decision trees as an alternative
  to artificial neural networks for particle identification}'',} \textit{ Nucl.
  Instrum. Meth.} \textbf{ 543} (2005) 577,
  \href{http://dx.doi.org/10.1016/j.nima.2004.12.018}{\doi{10.1016/j.nima.2004.12.018}}.

\bibitem{Verkerke:622147}
\hrefCMSnoop {}{W.~Verkerke and D.~P. Kirkby, ``{The RooFit toolkit for data
  modeling}'',} in \textit{ {Computing in High Energy and Nuclear Physics
  (CHEP03): Proceedings, La Jolla, USA, March, 2003}}, pp.~{Note: eConf
  C030324,186 (2003) MOLT007}.
\newblock 2003.
\newblock
\href{http://www.arXiv.org/abs/physics/0306116}{\texttt{arXiv:physics/0306116}}.
\newblock

\bibitem{DAgostini:1995jqe}
\hrefCMSnoop {}{G.~D'Agostini, ``{Probability and measurement uncertainty in
  physics: A Bayesian primer}'',} (1995).
\href{http://www.arXiv.org/abs/hep-ph/9512295}{\texttt{arXiv:hep-ph/9512295}}.

\bibitem{BAER1990127}
\hrefCMSnoop {}{H.~Baer, J.~Ohnemus, and J.~F. Owens, ``A calculation of the
  direct photon plus jet cross section in the next-to-leading-logarithm
  approximation'',} \textit{ Phys. Lett. B} \textbf{ 234} (1990) 127,
  \href{http://dx.doi.org/10.1016/0370-2693(90)92015-B}{\doi{10.1016/0370-2693(90)92015-B}}.

\bibitem{PhysRevD.42.61}
\hrefCMSnoop {}{H.~Baer, J.~Ohnemus, and J.~F. Owens,
  ``Next-to-leading-logarithm calculation of direct photon production'',}
  \textit{ Phys. Rev. D} \textbf{ 42} (1990) 61,
  \href{http://dx.doi.org/10.1103/PhysRevD.42.61}{\doi{10.1103/PhysRevD.42.61}}.

\bibitem{Accardi:2016qay}
A.~Accardi\hrefCMSnoop {}{ {et~al.}, ``{Constraints on large-$x$ parton
  distributions from new weak boson production and deep-inelastic scattering
  data}'',} \textit{ Phys. Rev. D} \textbf{ 93} (2016) 114017,
  \href{http://dx.doi.org/10.1103/PhysRevD.93.114017}{\doi{10.1103/PhysRevD.93.114017}},
\href{http://www.arXiv.org/abs/1602.03154}{\texttt{arXiv:1602.03154}}.

\bibitem{Bourhis:1997yu}
\hrefCMSnoop {}{L.~Bourhis, M.~Fontannaz, and J.~P. Guillet, ``{Quarks and
  gluon fragmentation functions into photons}'',} \textit{ Eur. Phys. J. C}
  \textbf{ {2}} (1998) 529,
  \href{http://dx.doi.org/10.1007/s100520050158}{\doi{10.1007/s100520050158}},
\href{http://www.arXiv.org/abs/hep-ph/9704447}{\texttt{arXiv:hep-ph/9704447}}.

\end{thebibliography}\endgroup

\cleardoublepage \appendix\section{The CMS Collaboration \label{app:collab}}\begin{sloppypar}\hyphenpenalty=5000\widowpenalty=500\clubpenalty=5000\vskip\cmsinstskip
\textbf{Yerevan Physics Institute, Yerevan, Armenia}\\*[0pt]
A.M.~Sirunyan, A.~Tumasyan
\vskip\cmsinstskip
\textbf{Institut f\"{u}r Hochenergiephysik, Wien, Austria}\\*[0pt]
W.~Adam, F.~Ambrogi, E.~Asilar, T.~Bergauer, J.~Brandstetter, M.~Dragicevic, J.~Er\"{o}, A.~Escalante~Del~Valle, M.~Flechl, R.~Fr\"{u}hwirth\cmsAuthorMark{1}, V.M.~Ghete, J.~Hrubec, M.~Jeitler\cmsAuthorMark{1}, N.~Krammer, I.~Kr\"{a}tschmer, D.~Liko, T.~Madlener, I.~Mikulec, N.~Rad, H.~Rohringer, J.~Schieck\cmsAuthorMark{1}, R.~Sch\"{o}fbeck, M.~Spanring, D.~Spitzbart, W.~Waltenberger, J.~Wittmann, C.-E.~Wulz\cmsAuthorMark{1}, M.~Zarucki
% ...
\vskip\cmsinstskip
% ...
1:  Also at Vienna University of Technology, Vienna, Austria\\
% ...
\end{sloppypar}
\end{document}